\documentclass[reprint,nofootinbib,...]{revtex4-1} 
\usepackage{amsmath}%
\usepackage{amsthm,amssymb}
\usepackage{graphicx}
\usepackage{epstopdf}
\usepackage{xcolor}
\usepackage{algcompatible}
\usepackage[size=small]{caption}
\usepackage{etoolbox}
\usepackage{booktabs}
\usepackage{multirow}
\usepackage[utf8]{inputenc}
\usepackage[colorinlistoftodos]{todonotes}

\usepackage[matrix,frame,arrow]{xypic}
\usepackage[braket,qm]{qcircuit}

\AtBeginEnvironment{algorithm}{\noindent\hrulefill\par\nobreak\vskip-5pt}
\usepackage{newfloat}
\usepackage{booktabs,makecell}

\DeclareFloatingEnvironment[
    fileext=loa,
    listname=List of Algorithms,
    name=ALGORITHM,
    placement=tbhp,
]{algorithm}
\DeclareCaptionFormat{algorithms}{\vskip-15pt\hrulefill\par#1#2#3\vskip-6pt\hrulefill}
\algrenewcommand{\algorithmiccomment}[1]{\hskip3em// {\it #1}}

\usepackage[hyperfootnotes=true]{hyperref}
\hypersetup{
 colorlinks=true,
 citecolor=blue,
 linkcolor=blue,
 urlcolor=blue}

\usepackage{comment}
\usepackage[matrix,frame,arrow]{xypic}
\usepackage[braket,qm]{qcircuit}

\def\df{{\rm d}}

\begin{document}

\title{A quantum algorithm for high energy physics simulations}

\author{Benjamin Nachman}
\email{bpnachman@lbl.gov}

\affiliation{Physics Division, Lawrence Berkeley National Laboratory, Berkeley, CA 94720, USA}

\author{Davide Provasoli}
\email{davideprovasoli@lbl.gov}

\affiliation{Physics Division, Lawrence Berkeley National Laboratory, Berkeley, CA 94720, USA}

\author{Wibe A. de Jong}
\email{WAdeJong@lbl.gov}

\affiliation{Computational Research Division, Lawrence Berkeley National Laboratory, Berkeley, CA 94720, USA}

\author{Christian W. Bauer}
\email{cwbauer@lbl.gov}

\affiliation{Physics Division, Lawrence Berkeley National Laboratory, Berkeley, CA 94720, USA}

\begin{abstract}
Simulating quantum field theories is a flagship application of quantum computing. However, calculating experimentally relevant high energy scattering amplitudes entirely on a quantum computer is prohibitively difficult. 
It is well known that such high energy scattering processes can be factored into pieces that can be computed using well established perturbative techniques, and pieces which currently have to be simulated using classical Markov Chain (MC) algorithms. These classical MC simulation approaches work well to capture many of the salient features, but cannot capture all quantum effects. To exploit quantum resources in the most efficient way, we introduce a new paradigm for quantum algorithms in field theories.  This approach uses quantum computers only for those parts of the problem which are not computable using existing techniques. In particular, we develop a polynomial time quantum final state shower that accurately models the effects of intermediate spin states similar to those present in high energy electroweak showers. The algorithm is explicitly demonstrated for a simplified quantum field theory on a quantum computer. 
\end{abstract}

\date{\today}
\maketitle


While quantum computers hold great promise for efficiently solving classical problems such as querying databases~\cite{Grover:1996rk} or factoring integers into primes~\cite{Shor:1994jg}, their most natural application is to describe inherently quantum physical systems~\cite{Feynman1982}.   The most direct connection between quantum systems and quantum computers occurs for analog circuits that try to mimic the evolution of a Hamiltonian as closely as possible~\cite{Georgescu:2013oza}.  
A promising alternative to analog circuits are digital quantum circuits, which use quantum algorithms to describe inherently quantum physical systems without directly implementing the system's Hamiltonian.  However, many physical systems are too complex to fully model with a quantum circuit using near-future noisy intermediate scale devices~\cite{Preskill2018quantumcomputingin}. This is true for a generic quantum field theory.  While tools have been developed to model quantum field theories by discretizing spacetime~\cite{Jordan:2011ne} and even including continuous quantum numbers~\cite{Marshall:2015mna}, the number of quantum bits required to compute any relevant scattering amplitude is impractically large.  Results with simplified quantum field theories on a lattice are promising~\cite{Martinez:2016yna}, but the full dynamics of high energy scattering processes are too complex for both lattice methods as well as traditional perturbation theory when the number of final state particles becomes too large.  A promising new avenue for quantum algorithms for quantum field theories is to use them only for the parts of the calculation that are computationally intractable using standard techniques.

It is well known that factorization theorems can be derived that separate shorter distance from longer distance physics~~\cite{Collins:1987pm,Sterman:1995fz,Jaffe:1996zw,Bauer:2002nz}. The shortest distance physics of the hard-scattering process is the most difficult part to simulate using lattice techniques (either using classical or quantum computers) and can in most cases be computed reliably using perturbative techniques.  The longest distances correspond to hadronization effects for which often non-perturbative models are used to describe the physics.  In between these two scales, one has to describe radiation occurring with soft and collinear divergences, which requires techniques beyond standard perturbation theory. A successful classical approach for simulating the dynamics of this final state radiation is known as the parton shower~\cite{PhysRevD.98.030001}, which relies on reorganizing the traditional perturbative series about a fundamental coupling constant to instead expand around the collinear and soft limit of emissions~\cite{Sterman:1977wj,Almeida:2014uva}. This leads to different series expansions where each term in the new series includes infinitely many terms from the original series expansion and is the basis of parton shower Monte Carlo (MC) programs~\cite{Sjostrand:2006za,Bahr:2008pv,Gleisberg:2008ta,Buckley:2011ms}, which are a key component of high energy quark and gluon scattering simulations.
Parton shower models are implemented using classical Markov Chain MC (MCMC) algorithms to efficiently generate high multiplicity radiation patterns.  This reliance on classical MCMC algorithms implies that several quantum interference effects need to be neglected.  While many current analyses are nearly insensitive to such effects, future work will analyze the final state radiation in more detail and new studies will be enabled with calculations that include new quantum effects.

 Our goal is to develop a quantum circuit describing the quantum properties of parton showers.  In this work, we consider showers with quantum interferences from different intermediate particles, using a simplified model that captures these effects without having to introduce the full complexity of the Standard Model (SM).   The variable describing the scale of the shower evolution is discretized and at each step an emission can occur or not.   We will show that a classical MCMC is not able to capture the important quantum interference effects in this model, and that a full classical calculation scales exponentially with the number of steps\footnote{There are efficient algorithms to account for spin correlations in quantum chromodynamics~\cite{Knowles:1988vs,Knowles:1988hu,Collins:1989gx,Richardson:2001df}, but these do not apply to our model or more generally to any model such as $SU(2)$ where the emission probability depends on the spin~~\cite{Richardson:2018pvo}.}.  The proposed quantum algorithm will be able to sample from the full probability distribution in polynomial time.  

To begin, consider a simple quantum field theory, with two types of fermion fields, $f_{1}$ and $f_{2}$, interacting with one scalar boson $\phi$ governed by the following Lagrangian:
\begin{align}
\label{Lagrangian}\nonumber
{\cal L} = & \bar f_{1} (i\slash\!\!\!{\partial} + m_1) f_{1} + \bar f_{2} (i \slash\!\!\!{\partial} + m_2) f_{2} + (\partial_\mu \phi)^2  \\
&\hspace{1mm}
+ g_{1}  \bar f_{1} f_{1} \phi 
+ g_2  \bar f_{2} f_{2} \phi  + g_{12} \left[ \bar f_{1} f_{2} + \bar f_{2} f_{1} \right]\phi 
\,.
\end{align}
The first three terms in Eq.~\ref{Lagrangian} describe the kinematics of the fermions and scalar while the latter three terms govern their interactions.  In particular, the collinear dynamics of the theory are that the fermions can radiate scalars ($f_i\rightarrow f_j\phi$) and scalars can split into fermion pairs ($\phi\rightarrow f_i\bar{f}_j$).  These couplings of fermions to scalar bosons occur in the Higgs sector of the SM, and it has been demonstrated that the final state collinear radiation at high energy can be written in terms of a parton shower~\cite{Chen:2016wkt,Bauer:2018xag}. This model can contain important quantum interference effects when all couplings are non-zero, since the unobserved intermediate state of the fermions can be a superposition of $f_i$ for $i\in\{1,2\}$.  

In the limit $g_{12} \to 0$ one can derive an efficient MCMC method for calculating high-multiplicity cross sections.  This is performed by introducing four splitting functions, two for a fermion radiating a scalar $(P_{i \to i \phi}(\theta) = g_i^2 \hat P_f(\theta))$ and two for the scalar splitting into fermions $(P_{\phi\to i i}(\theta) = g_i^2 \hat P_\phi(\theta))$, where $\theta$ is the scale at which the splitting occurs and $\hat{P}(\theta)$ encodes the energy scale-dependence of the emission probability.  There are many formally equivalent definitions of the scale; here we use a common choice: the opening angle of the emission with respect to the emitter.  In addition to the splitting functions, another important quantity is the no-branching probability (Sudakov factor):
\begin{align}
\Delta_{i,k}(\theta_1, \theta_2) &= \exp\left[ -g_{i}^2 \int_{\theta_1}^{\theta_2} \df \theta' \hat P_k(\theta')\right]
\,.
\end{align}
The Sudakov factor encapsulates the virtual (and unresolved real) contributions and is responsible for the reorganization of the perturbation series (`resummation') mentioned above. The Sudakov factor and splitting function satisfy the unitarity relation
\begin{align}
\Delta_{i,k}(\theta_1, \theta_2) + g_i^2\int_{\theta_1}^{\theta_2} \! \df \theta \,\hat P_{k}(\theta)\, \Delta_{i,k}(\theta_1, \theta) = 1
\,.
\end{align}
A classical parton shower would then efficiently sample from the cross section using a Markov Chain algorithm by generating one emission at a time, conditioned on the last emission.  In particular, at a given step $n$ in $\theta$, there are at most $n$ particles and the probability that none of them radiate or split is $\prod_{j=1}^N\Delta_{i_j,k_j}$.  If something does happen at a given step, the probabilities are proportional to the appropriate splitting function.

When $g_{12}>0$, there are now multiple histories with unmeasured intermediate fermion types which contribute to the same final state.  Therefore, the above MCMC is invalid because one must include all possible histories and cannot condition on a given state.  Including all of the interference effects requires accounting for all histories at the amplitude level and only computing probabilities at the end of the evolution.  When the $g_{12}\ll 1$, the evolution is dominated by a single emission, which can be properly treated using a density matrix formalism~\cite{Chen:2016wkt}, where each splitting function is represented through a splitting matrix.  For example, the fermion splitting matrix is $P_{i\rightarrow j\phi}(\theta) \ket{f_{i}} \bra{f_{j}}$ (outer product of a ket and bra gives a matrix).  When there is more than one emission during the evolution, this matrix formalism is insufficient and one must compute the full amplitude for which there are $\mathcal{O}(2^N)$ possible histories [see App.~\ref{sec:methods}].

We propose an efficient solution by keeping track of amplitudes and not probabilities using  quantum computer.  A quantum circuit implementing the quantum final state radiation algorithm for one of $N$ steps is given by the following diagram:
\[
\Qcircuit @C=0.5em @R=0.8em @!R{
\lstick{\ket{p}} & {/} \qw &  \gate{R^{(m)}} &   \measure{\mbox{$p$}} \qwx[3]  & \qw  & \measure{\mbox{$p$}} \qwx[1]  &  \sgate{U^{(m)}_p}{1} &  \gate{{R^{(m)}}^{\dagger}}  & \qw   \\
\lstick{\ket{h}} & {/} \qw  & \qw & \qw & \qw & \sgate{U_{h}}{1}   &   \measure{\mbox{$h$}} \qwx[-1]  & \qw & \qw  \\ 
\lstick{\ket{e}} & \qw  & \qw & \qw & \gate{U^{(m)}_{e}} &  \measure{\mbox{$e$}} \qwx[1]   & \qw & \qw     & \qw   \\
\lstick{\ket{n_{\phi}}} & {/} \qw & \qw  & \multigate{2}{U_{\rm count}} & \measure{\mbox{$n_\phi$}} \qwx[-1]& \multigate{2}{U_{h}} & \qw  & \qw  & \qw     \\
\lstick{\ket{n_a}} & {/} \qw  & \qw &  \ghost{U_{\rm count}} & \measure{\mbox{$n_a$}} \qwx[-1]  &  \ghost{U_{h}} &  \qw & \qw  & \qw  \\ 
\lstick{\ket{n_b}} & {/} \qw  & \qw & \ghost{U_{\rm count}}  &\measure{\mbox{$n_b$}} \qwx[-1] & \ghost{U_{h}} & \qw & \qw   & \qw  
}
\]

\vspace{2mm}

The circuit calls for six registers, which are are detailed in the App.~\ref{sec:methods} and summarized in Tables~\ref{tab:registers} and~\ref{tab:complexity}. The initial state of $\ket{p}$ consists of $n_I$ particles (which can be fermions or bosons) in the $f_{1/2}$ basis. One starts by rotating this initial particle state from the $f_{1/2}$ basis to a diagonal $f_{a/b}$ basis, using a simple unitary $R^{(m)}$ operation discussed in App.~\ref{sec:methods}. Then, a series of operations evolving the particles states are applied: the number of particles of each type are counted ($U_\text{count}$), Sudakov factors are used to determine if an emission occurred ($U_e^{(m)}$), given an emission, a particular particle is chosen to radiate/branch ($U_h$), and the resulting particle state is updated ($U_p^{(m)}$).  Finally, the state is rotated back to the $f_{1/2}$ basis through the ${R^{(m)}}^{\dagger}$ operation. This process is repeated for all of the $N$ steps.  

Performing the evolution in the $f_{a/b}$ basis and then rotating to the $f_{1/2}$ basis, creates interferences between equivalent final states which had different intermediate fermions. One event is generated by measuring all of the qubits after the final rotation back to the $f_{1/2}$ basis.  By repeating the entire process, we can generate a large number of events which we can then use to compute physical observables for our theory.  As discussed in App.~\ref{sec:methods}, the algorithm presented can be simplified significantly if the a subset of qubits representing the history register $\ket{h}$ can be measured at the end of each step. This fixes the total number of particles as well as the total number of bosons. The number of standard quantum gates (single qubit and CNOT gates) required at each step is discussed in App.~\ref{sec:methods} and summarized in Table~\ref{tab:complexity} with and without the repeated measuring of the history register. Comparing the scaling of the quantum algorithm with an efficient classical algorithm, discussed in App.~\ref{sec:methods}, the quantum algorithm outperforms the classical algorithm once the number of emitted particles exceeds ${\cal O}(10)$, if the history register is measured after each step. Without this repeated measurement, the number of steps required for the quantum algorithm to beat the classical one depends on the size of the coupling constants $g_{1,2}$ as the classical scaling goes with the number of fermions and not the (much) larger number of steps.

The practical challenge with above circuit is that it requires more connected qubits and operations than are currently available in state-of-the-art hardware.  In order to show an implementation of our algorithm, we therefore consider a special case that is amenable to measurement on existing technology.  This special case ignores the $\phi\rightarrow f\bar{f}$ splitting (naturally suppressed in gauge theories, but not in the scalar-only theory), ignores the running coupling, and has only a single fermion (possibly in a superposition) as the initial state. This results in a much simpler circuit since there is only one fermion, but an arbitrary number of scalars.  A decomposition of the resulting circuit into single qubit and CNOT gates requires $n_{{\rm gates}} = 12 \, N + 2$ (see App.~\ref{sec:methods}).  This model is however still sufficiently complex that the classical MCMC described earlier\footnote{While the standard parton shower-inspired MCMC algorithm fails, we have discovered a quantum-inspired classical algorithm that can efficiently sample from the full probability distribution~~\cite{Provasoli}.  However, this algorithm only works when neglecting the $\phi\rightarrow f\bar{f}$ and cannot solve our full model.} fails to capture important quantum effects when $g_{12}\neq 0$.

\begin{figure*}
\centering
\includegraphics[width=0.5\textwidth]{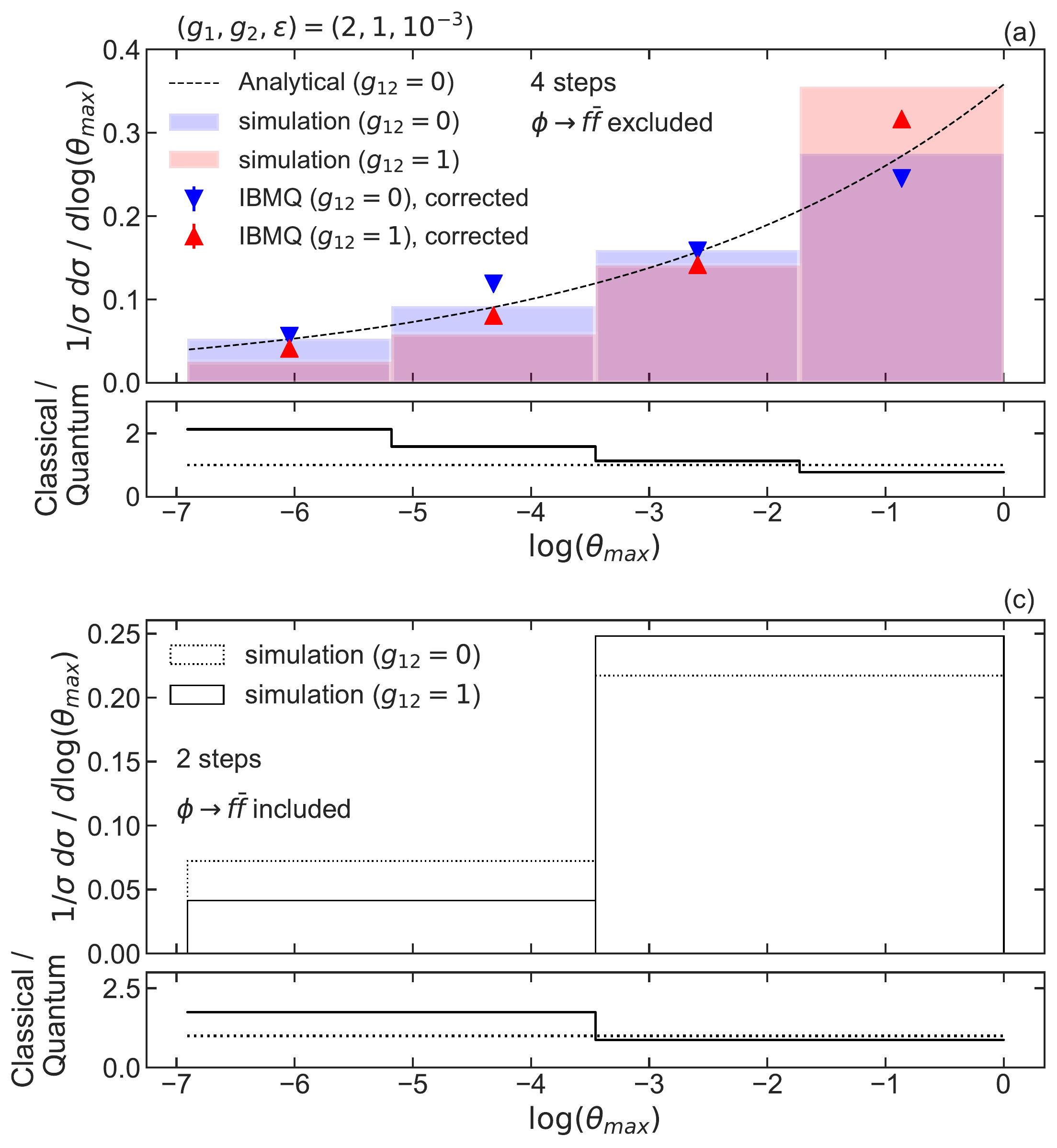}\includegraphics[width=0.5\textwidth]{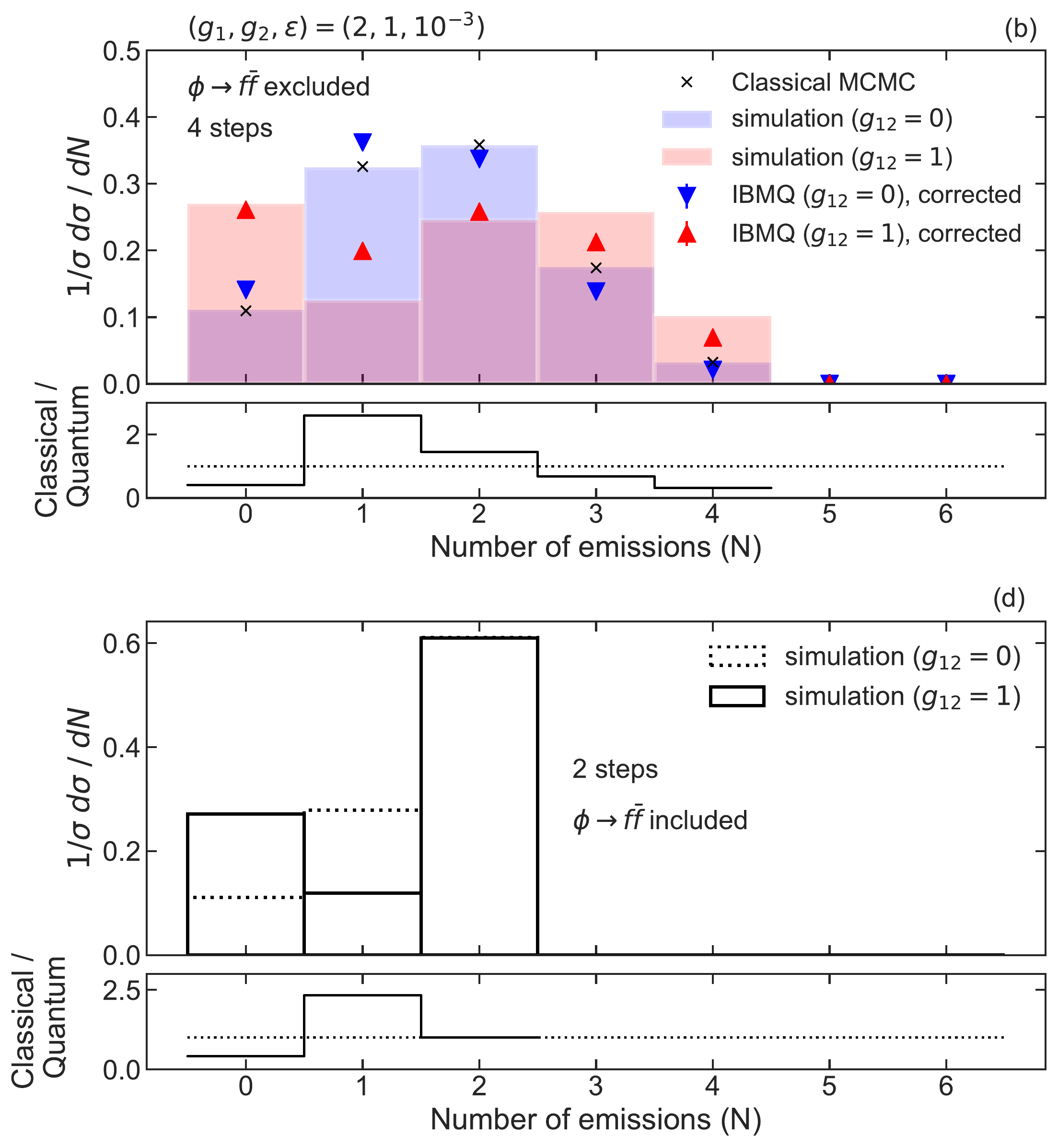}
\caption{The normalized differential cross section for $\log\theta_{\rm max}$ (a,c) and the number of emissions (b,d). Interference effects are turned on ($g_{12}=1$) and off ($g_{12}=0$), where the classical simulations/calculations are expected to agree with the quantum simulations and measurements.  The top plots (a,b) show results for the case where $\phi\rightarrow f\bar{f}$ is excluded as this can be run on current quantum hardware.  The bottom plots (c,d) include the $\phi\rightarrow f\bar{f}$ with fewer steps to reduce the computational complexity.  The ratio plots compare the $g_{12}=0$ and $g_{12}=1$ simulation.  Over $10^5$ events contribute to each line and the statistical uncertainties are therefore negligible.  Quantum measurements are corrected for readout errors, as described in App.~\ref{sec:methods}.}
\label{fig:quantum}
\end{figure*}

Figure~\ref{fig:quantum} presents the normalized differential cross sections for the logarithm of the largest emission angle (a,c) as well as the number of emissions (b,c) for both classical simulations/calculations, quantum simulators~\cite{Qiskit}, and chip experiments of public and Q Hub member quantum chips through cloud access on the IBM Quantum Experience. All cases are started from the initial state containing a single $f_1$ fermion. The data of experimental measurements shown in Figure~\ref{fig:quantum} were collected on the IBM Q Johannesburg chip. This quantum computer has twenty qubits, and to restrict the gate depth and hardware fidelity challenges we choose to simulate $N=4$ steps.  The 4-step circuit on 5 qubits requires 48 gate operations, of which 17 are 2-qubit operations. Details of the experiments, including measurement corrections are discussed in App.~\ref{sec:methods}. In addition to presenting the simplified model with both quantum hardware and simulations, Figure~\ref{fig:quantum} also shows a simulation with the full model (including $\phi\rightarrow f\bar{f}$) for 2 steps.  

 \begin{table}
\centering
\begin{tabular}{ccc}
\toprule
\thead{Register} & \thead{Purpose} & \thead{\# of qubits} \\
\midrule
$\ket{p}$ & Particle state & $ 3  (N  + n_I )$ \\
$\ket{h}$ & Emission history & $N \lceil \log_2(N  + n_I  ) \rceil $ \\
$\ket{e}$ & Did emission happen? & $1$ \\
$\ket{n_{\phi}}$ & Number of bosons & $ \lceil \log_2(N  + n_I ) \rceil $ \\
$\ket{n_a}$ & Number of $f_a$ & $ \lceil \log_2(N  + n_I ) \rceil $ \\
$\ket{n_b}$ & Number of $f_b$ & $ \lceil \log_2(N  + n_I ) \rceil $ \\
\bottomrule
\end{tabular}
\caption{All of the registers in the quantum circuit with the number of qubits they require for $N$ steps and $n_I$ initial particles.  The symbol $\lceil \ldots \rceil$ denotes the ceiling function.}
\label{tab:registers}
\end{table}

\begin{table}
\centering
\begin{tabular}{cccc}
\toprule
\multirow{2}*{\thead{Operation}} 
& \multicolumn{2}{c}{\thead{Scaling}} & \multicolumn{1}{c}{\thead{\#  gates \\ (default alg.)}} \\
&\thead{default algorithm}   & \thead{measure $\ket{h}$} & \thead{N = 4}   \\
\midrule
\thead{count particles [$U_{\rm count}]$} &$N \ln N$  &$N \ln n_f$ & $4.93 \times10^2$\\
\thead{decide emission  [$U_e$]} & $N^4 \ln N$  &$N n_f \ln n_f$ & $9.29 \times10^3$  \\
\thead{create history  [$U_h$]} & $N^5 \ln N$  &$N n_f^2\ln n_f$ & $1.96 \times10^5$ \\
\thead{adjust particles [ $U_p$]} &$N^2 \ln N$  &$N n_f \ln n_f$ & $5.01 \times10^3$\\
\hline
\thead{classical algorithm} & \multicolumn{2}{c}{$N2^{n_f/2}$} \\
\bottomrule
\end{tabular}
\caption{List of the circuit operations with the number of standard gates required for given numbers of steps assuming $n_I = 1$.  Further details about the calculations involved and the counting of the number of gates can be found in App.~\ref{sec:methods}.  The third column provides the scaling assuming that classical registers could be used to store the history qubit at each step.  This is not implemented in the algorithm shown in Fig.~\ref{fig:quantum}, but may be possible on near-term hardware.}
\label{tab:complexity}
\end{table}

When interference effects are turned off ($g_{12}=0$), we find excellent agreement for all observables between both the classical and quantum simulator results as well as the quantum computer measurements. For $g_{12} = 1$ the spectra are shifted to the right, leading to more emissions and at larger angles. For all quantum simulations the fraction of events with no emissions (first bin in (b) and (d)) agree separately for each value of $g_{12}$. This is because the simulation is started with a single fermion state, where the splitting $\phi \to f \bar f$ is irrelevant. For a higher number of emissions, the $\phi \to f \bar f$ splitting affects the distribution, and in particular lowers the fraction of events with a single emission. 

The experimental data points obtained running the 48 operation simulation on the IBM Q Johannesburg quantum computer are in agreement with the quantum simulator results, clearly showing the role of interference when the interaction is turned on (from $g_{12}=0$ to $g_{12}=1$). Some differences can be observed between the quantum simulator and the actual quantum computer experiment, which can be attributed to the noise present in the existing hardware. These results are an important and first-of-its-kind experimental proof-of-principle for the algorithm and are the first step in the realization of a larger and more precise calculations on future hardware.


Future extensions of the algorithm can bring it closer to the SM.   To include the full three-dimensional kinematics of parton splittings, one would need to augment the procedure to sample momentum fractions and azimuthal angles.  It may be possible to achieve this with a hybrid quantum-classical approach as the probability distribution factorizes.  The current model includes all of the discrete quantum numbers of the Higgs sector of the SM with one fermion.  Adding additional fermions is algorithmically simple and is a linear computational cost.  A further extension to the $SU(2)$ component of the SM would be able to use nearly the same algorithm, but with more fermions and additional bosons to represent the $W^\pm$ and $Z$ bosons.  While the numerical implications of such electroweak showers with multiple bosons may be small at previous and current colliders, such calculations may be important for future colliders as well as indirect dark matter search experiments with ultra high-energy particles~~\cite{Blanco:2017sbc} with a Higgs or full $SU(2)$-like structure.  Further extensions to the strong force $SU(3)$ may be possible, but are more complicated because of the important role of low energy and not just collinear radiation.  There is also an extensive literature on approximate amplitude-based approaches using classical methods~~\cite{Nagy:2007ty,Nagy:2019pjp,Platzer:2018pmd,Platzer:2012np,Martinez:2018ffw,Forshaw:2019ver,Isaacson:2018zdi}, which may also have potential synergy with quantum algorithms in the future.

With improved quantum hardware beyond the current noisy intermediate scale devices~\cite{Preskill2018quantumcomputingin}, our algorithms will be able to produce calculations that are currently not possible with classical devices.  The richness of quantum phenomena in high energy physics makes them an excellent testbed for studying the power of quantum algorithms.  By focusing on final state radiation, quantum algorithms may be able to provide key insight into the dynamics of quantum field theories underlying the laws of nature.

\section*{ACKNOWLEDGMENTS}

We thank A. Ryd for useful discussions on the treatment of spin correlations in classical MCMC programs.  This work is supported by the DOE under contract DE-AC02-05CH11231. In particular, support comes from Quantum Information Science Enabled Discovery (QuantISED) for High Energy Physics (KA2401032).  We acknowledge access to quantum chips and simulators through the IBM Quantum Experience and Q Hub Network.

\appendix

\section{Methods}
\label{sec:methods}

\subsection{The registers of the quantum circuit}
\label{sec:registers}
The quantum circuit introduced in this paper has a total of 6 registers. The first register, $\ket{p}$,  contains the flavour information about each particle. Each particle in the system can be in one of 6 states $\ket{0}$, $\ket{\phi}$, $\ket{f_{a/b}}$, and $\ket{\bar f_{a/b}}$.  To encode these $6$ states one requires 3 qubits, and we choose the representation as
\begin{align}
\label{eq:states}
\ket{p}_i = \left( \begin{array}{cc}000 \\ 001 \\ 010 \\ 011 \\ 100 \\ 101 \\ 110 \\ 111 \end{array} \right) = \left( \begin{array}{cc} 0 \\ \phi \\ -  \\ -  \\ f_1 / f_a \\ f_2 / f_b \\ \bar{f}_1 / \bar{f}_a \\ \bar{f}_2 / \bar{f}_b \end{array} \right) \,
\,,
\end{align}
where the third and fourth states are not used and one chooses $f_{1/2}$ and $f_{a/b}$ before and after the basis change discussed in the next section. Since there can be up to $N+n_I$ particles in the system (where $n_I$ is the initial number of particles and $N$ is the number of steps), one needs a total of 
\begin{align}
\dim[\ket{p}] =3(N+n_I)
\end{align}
qubits to encode this register.

The second register, $\ket{h}$, holds the information about which particle emitted a particle at a given step. At the start of the $m^\text{th}$ step (where the first step has $m = 0$), there are up to $m+n_I $ particles that can have emitted the extra particle, and at the $m^\text{th}$ step $\ket{h}_m$ needs to be able to hold the integers $0 \ldots m+n_I$ (where $0$ denotes no particle having emitted something). When considering $N$ steps, the register therefore needs to hold $\sum_{m=0}^{N-1} (m+n_I)  = N(N+2n_I+1)/2$ integers, requiring 
\begin{align}
\dim[\ket{h}] = \lceil \log_2[N(N+2n_I+1)/2] \rceil
\,,\end{align}
where $\lceil \ldots \rceil$ denotes the ceiling function. It might be simpler to have each $\ket{h}_m$ be of the same size, in which case each $\ket{h}_m$ would need to hold the integers $0 \ldots N+n_I-1$. This would require 
\begin{align}
\dim[\ket{h}] = N \lceil \log_2[(N+n_I)] \rceil
\end{align}
qubits. 

The third register, $\ket{e}$ temporarily holds the information whether an emission has occurred in the current step. This is binary information, and therefore requires a single qubit, giving 
\begin{align}
\dim[\ket{e}] = 1
\,.
\end{align}

The remaining three registers are count registers, which temporarily hold the information about how many bosons, fermions of type $a$ and fermions of type $b$ (counting both $f$ and $\bar f$) are in the current state. Since the count registers are used for every step, they have to hold the integers $0, \ldots, N+n_I$. We again choose the binary representation to hold these integers, and one needs 
\begin{align}
\dim[\ket{n_\phi}] = \dim[\ket{n_{a/b}}] =  \lceil \log_2[(N+n_I)] \rceil
\end{align}
 qubits. 

At the start of the circuit, all work registers $\ket{e}$, $\ket{n_\phi}$, $\ket{n_a}$, and $\ket{n_b}$ are initialized to $\ket{0}$, where for the count registers $\ket{0}$ refers to the integer $0$ in binary notation. For the physical registers, all history registers $\ket{h}_m$ as well as the particle registers $\ket{p}_{m>n_I}$ are initialized to zero. The only non-zero registers are $\ket{p}_{m\leq n_I}$, which are initialized to the initial particle content (possibly in a superposition). 

\subsection{Diagonalizing the splitting matrix}
\label{sec:diagonalize}
The splitting matrix can be written in terms of the coupling constants $g_1$, $g_2$ and $g_{12}$ as
\begin{align}
P_{i\rightarrow j\phi}(\theta) = G_{ij} \, \hat P(\theta) \equiv 
\left( \begin{array}{cc} g_1 & g_{12} \\ g_{12} & g_2 \end{array} \right)
\hat P(\theta)
\,.
\end{align}
The coupling matrix $G$ can be diagonalized as
\begin{align}
\label{eq:diagonalization}
G^{\rm diag} = U G U^\dagger = \left( \begin{array}{cc} g_a & 0 \\ 0 & g_b \end{array} \right)\,,
\end{align}
with
\begin{align}
g_a = \frac{g_1 + g_2 - g'}{2}\,, \qquad g_b= \frac{g_1 + g_2 + g'}{2}
\,,
\end{align}
where
\begin{align}
g' = {\text {sign}(g_2 - g_1)}\sqrt{(g_1 - g_2)^2 + 4 g_{12}^2}
\,.
\end{align}
The matrix $U$ in Eq.~\eqref{eq:diagonalization} is given by
\begin{align}
\label{eq:UfDef}
U = \left( \begin{array}{cc} \sqrt{1-u^2} & u \\ -u & \sqrt{1-u^2}\end{array} \right)\,,
\end{align}
with
\begin{align}
u = \sqrt{\frac{(g_1 - g_2 + g')}{2 g'}}
\,.
\end{align}

Since each particle is represented by a 3-qubit state, the operation $R$ that rotates a single particle from the $f_{1/2}$ basis to the $f_{a/b}$ basis is represented by a $8 \times 8$ unitary matrix $R$, and it must be applied to all of these 3-qubit particle states. It is defined in terms of the matrix $U$, introduced in Eq.~(\ref{eq:UfDef}). For the representation of the particles given in the previous section, one has

\begin{align}
\label{eq:R}
R = \left( \begin{array}{cccc} I & 0 & 0 & 0  \\   0& I & 0& 0   \\  0 &0 & U & 0 \\ 0 & 0 &0 & U\\ \end{array} \right)\,,
\end{align}

\noindent where $I$ denotes the $2 \times 2$ identity matrix. The rotation $R$ correctly mixes the fermion states, while it leaves alone the $\ket{\phi}$ and $\ket{0}$ states. Because of the running of the coupling constants the matrix $U$, and in turn the matrix $R$, will be different at each step in the evolution. 

\subsection{Populating the register for counting the particles}
\label{sec:counting}
As discussed, at the beginning of each step the count registers $\ket{n_{\phi}}$, $\ket{n_a}$ and $\ket{n_b}$ are in the state $\ket{0}$. To perform this counting we apply the controlled $U^{(m)}_{\rm count}$ gate, which is given by
\[
\Qcircuit @C=0.8em @R=.5em @!R{
\lstick{\ket{p}} & {/} \qw & \measure{\mbox{$\phi$}} \qwx[2]    & \measure{\mbox{$a$}} \qwx[3]     & \measure{\mbox{$b$}} \qwx[4]   & \qw   &&& &&&\lstick{\ket{p}} & {/} \qw &\measure{\mbox{$p$}}  \qwx[2] &\qw   \\
 &  &  &  &  &  &  & \\
\lstick{\ket{n_{\phi}}} & {/} \qw & \gate{U_+} & \qw & \qw & \qw &&\equiv&&&&\lstick{\ket{n_{\phi}}} &{/}\qw&\multigate{2}{U_{\rm count}} & \qw \\
\lstick{\ket{n_a}} & {/} \qw  &  \qw  &  \gate{U_+} & \qw & \qw&&&&&&\lstick{\ket{n_a}} &{/}\qw & \ghost{U_{\rm count}} & \qw \\ 
\lstick{\ket{n_b}} & {/} \qw & \qw & \qw &  \gate{U_+} & \qw &&&&&&\lstick{\ket{n_b}}  & {/}\qw&  \ghost{U_{\rm count}}& \qw
}
\]
For each particle in the state $\ket{p}$ we apply the unitary operation $U_+$ to the appropriate count register. The operation $U_+$ is defined on a set of integer states ranging from $0 \ldots N+n_I$ as 
\begin{align}
U_+ \ket{n} = \ket{n+1}_{{\rm mod}_{N+n_I}}
\,,
\end{align}
or in matrix form, $(U_+)_{ij}=1$ if $j=i+1$ mod $(N+n_I)$ and $0$ otherwise.
This is a simple operation, and the gate decomposition of the $U_+$ operator can be found when discussing the circuit decomposition.

\subsection{Sudakov factors in the quantum circuit}
\label{sec:sudakovinquantum}

In the $a/b$ basis the splitting can not change the flavour of the emitting fermion, and the evolution can therefore be described in terms of individual splitting functions and Sudakov factors, just as in a usual MCMC. For the fermions there are  2 different splitting functions
\begin{align}
P_{i \to i \phi}(\theta) &= g_i^2 \hat P_f(\theta)
\,,
\end{align}
where $i \in \{a,b\}$. The splitting of the bosons is given by
\begin{align}
P_{\phi \to i \bar i}(\theta) &= g_i^2 \hat P_\phi(\theta)
\,,
\end{align}
Using these splitting functions, one can define Sudakov factors, which describe the probability to have no emission from a given particle in a given step $m$. One finds
\begin{align}
\Delta_{i}(\theta_m) &\equiv \exp\left[ - \Delta \theta \, P_{i}(\theta_m)\right]
\nonumber\\
\Delta_{\phi}(\theta_m) &\equiv \exp\left[ - \Delta \theta \, P_{\phi}(\theta_m)\right] 
\,,
\end{align}
where
\begin{align}
P_{i}(\theta_m) &\equiv P_{i\to i \phi}(\theta_m)
\nonumber\\
P_\phi(\theta_m) &\equiv P_{\phi\to a \bar a}(\theta_m) + P_{\phi\to b \bar b}(\theta_m)
\,,
\end{align}
and
\begin{align}
\Delta \theta = \theta_m - \theta_{m+1}
\,.
\end{align}

The probability to have no emission from a state containing $n_\phi$ bosons and $n_{a/b}$ fermions of type $a/b$, is then given by
\begin{align}
\Delta^{(m)}(\theta_m) = \Delta^{n_{\phi}}_{\phi}(\theta_m) \Delta^{n_a}_a(\theta_m) \Delta^{n_b}_b(\theta_m)
\,.
\end{align}
From this one can derive the probability to have a branching at a given step, which is given by
\begin{align}
\qquad q_p(\theta_m) & \equiv \int_{\theta_m}^{\theta_{m+1}} d\theta P_p(\theta_m) \Delta_p(\theta_m, \theta)  
\nonumber\\
& =1 - \Delta_p(\theta_m, \theta_{m+1})
\,.
\end{align}
One therefore finds the unitarity condition
\begin{align}
\Delta_p(\theta_m, \theta_{m+1}) + q_p(\theta_m)   = 1
\,.
\end{align}

This splitting probability can be encoded in the quantum circuit through the rotation $U^{(m)}_e$ on the qubit $\ket{e}$. It starts off in the state $\ket{0}$ and is transformed to $\ket{1}$ if there is an emission and stays in the $\ket{0}$ state if there is no emission. 
The emission matrix is given by
\begin{align}
U^{(m)}_e = \left( \begin{array}{cc}\sqrt{\Delta^{(m)}(\theta_m) } & - \sqrt{1 -\Delta^{(m)}(\theta_m)} \\  \sqrt{1 - \Delta^{(m)}(\theta_m)} & \sqrt{\Delta^{(m)}(\theta_m)}  \end{array} \right)
\,.
\end{align}

\subsection{Selecting a particle to radiate or split}
\label{sec:whichemitted}

In order to select from which particle the emission occurred, one needs to ``loop'' over all particles in the register, up to the $(m+n_I)^\text{th}$ particle for step $m$, which can be written in terms of sub-operations
\[
\Qcircuit @C=0.5em @R=.5em @!R{
\lstick{\ket{p}_M} & {/} \qw  & \qw &  \qw   &\qw  & \qw & & & \measure{\mbox{$p$}} \qwx[4]  &  \qw & \qw     \\
&&&&&&\push{\rule{.1em}{0em}\ldots\rule{.1em}{0em}}&&& \\
\lstick{\ket{p}_2} & {/} \qw  & \qw &  \qw   & \measure{\mbox{$p$}} \qwx[2]  & \qw & & &\qw &  \qw& \qw      \\
\lstick{\ket{p}_1} & {/} \qw  &\measure{\mbox{$p$}} \qwx[1]  &  \qw   &\qw & \qw & & &\qw &  \qw& \qw      \\
\lstick{\ket{h}_m} & {/} \qw &  \sgate{U^{(m,1)}_{h}}{1}& \qw &  \sgate{U^{(m,2)}_{h}}{1}& \qw & &  & \sgate{U^{(m,M)}_{h}}{1}&\measure{\mbox{$\slash\!\!\!0$}} \qwx[1] & \qw  \\ 
\lstick{\ket{e}} & \qw &  \ctrl{1}  & \qw &  \ctrl{1}  & \qw &  \push{\rule{.1em}{0em}\ldots\rule{.1em}{0em}} & &   \ctrl{1}  & \gate{X}&\qw   \\
\lstick{\ket{n_{\phi}}} & {/} \qw & \multigate{2}{U^{(m,1)}_{h}} & \qw & \multigate{2}{U^{(m,2)}_{h}} & \qw & & & \multigate{2}{U^{(m,M)}_{h}} &  \qw& \qw      \\
\lstick{\ket{n_a}} & {/} \qw &   \ghost{U^{(m,1)}_{h}}    & \qw &   \ghost{U^{(m,2)}_{h}}    & \qw &  & &  \ghost{U^{(m,M)}_{h}}    & \qw& \qw    \\ 
\lstick{\ket{n_b}} & {/} \qw &  \ghost{U^{(m,1)}_{h}}  & \qw &  \ghost{U^{(m,2)}_{h}}  & \qw &  & & \ghost{U^{(m,M)}_{h}}  & \qw  & \qw  
}
\]

\vspace{-1mm}

Each sub-operation controls on one of the $\ket{p}_i$ in the particle register, and the final operation ensures that the emission qubit is back in the $\ket{0}$ state after the operation. The sub-operation which controls on the $k^\text{th}$ particle is given by
\[
\Qcircuit @C=1em @R=.5em @!R{
\lstick{\ket{p}_k}            & {/} \qw  &\measure{\mbox{$p$}} \qwx[1]               & \qw  & & & \measure{\mbox{$p$}} \qwx[1]   & \measure{\mbox{$\phi$}} \qwx[3]     &\measure{\mbox{$a$}} \qwx[4]    &  \measure{\mbox{$b$}} \qwx[5]   & \qw     \\
\lstick{\ket{h}_m}            & {/} \qw  & \sgate{U^{(m,k)}_{h}}{1}     & \qw & & &  \sgate{U^{(m,k)}_{h}}{1}& \qw  & \qw & \qw & \qw \\ 
\lstick{\ket{e}}            & \qw      & \ctrl{1}                               & \qw &  \push{\rule{.1em}{0em}\equiv\rule{.1em}{0em}} &  & \ctrl{1}  & \qw   & \qw  & \qw & \qw \\
\lstick{\ket{n_{\phi}}}  & {/} \qw & \multigate{2}{U^{(m,k)}_{h}} &  \qw & & & \measure{\mbox{$n_\phi$}} \qwx[1]   & \gate{U_-}    & \qw  & \qw  & \qw     \\
\lstick{\ket{n_a}}        & {/} \qw &   \ghost{U^{(m,k)}_{h}}         & \qw & & & \measure{\mbox{$n_a$}} \qwx[1]   & \qw  & \gate{U_-}& \qw   & \qw   \\ 
\lstick{\ket{n_b}}        & {/} \qw &   \ghost{U^{(m,k)}_{h}}         & \qw & && \measure{\mbox{$n_b$}} \qwx[-1]  & \qw & \qw & \gate{U_-}  & \qw   
}
\]

\vspace{-1mm}
\noindent Here a control on $\ket{p_k}$ means that the controlled unitary operation $U_h$ depends on the flavour of particle $p_k$, while $\phi_k$, $a_k$ and $b_k$ are true or false if $\ket{p}_k$ is either a boson, an $a$-fermion, or a $b$-fermion, respectively.

For each particle in $\ket{p}$ we apply $U^{(m,k)}_{h}$ if the emission has occurred in the given step. $U^{(m, k)}_{h}$ is a $2\times2$ unitary sub-matrix which always acts between the states $\ket{0}$ and $\ket{k}$ of $\ket{h}_m$. Defining
\begin{align}
P(n_{\phi},n_a,n_b)(\theta_m) = \sum_p n_p P_p(\theta_m)
\,,
\end{align}
where $P_a$, $P_b$ and $P_{\phi}$ are given coefficients, the mentioned $2\times 2$ submatrix is given by
\begin{align}
\label{eq:Uh}
U^{(m,k)}_h = \left( \begin{array}{cc}   \sqrt{1-\frac{P_{p_k}(\theta_m) }{P(n_{\phi}, n_a, n_b)}} & -\sqrt{\frac{P_{p_k}(\theta_m) }{P(n_{\phi}, n_a, n_b)}} \\\sqrt{\frac{P_{p_k}(\theta_m) }{P(n_{\phi}, n_a, n_b)}} & \sqrt{1-\frac{P_{p_k}(\theta_m) }{P(n_{\phi}, n_a, n_b)}} \end{array} \right)
\,.\end{align}
The coefficients of the matrix $U^{(m,k)}_h$ depend on $n_{\phi}$, $n_a$ and $n_b$, which is why we control on the count registers. Thus, if an emission has occurred, in each sub-operation the controlled $U^{(m,k)}_{h}$ gate rotates between the states $\ket{0}$ and $\ket{k}$ in the $\ket{h}$ register. This is done recursively in a way that builds up the correct amplitudes for each possible emission history. 

After each application of $U^{(m,k)}_h$, the count register is reduced, changing the value of $P(n_{\phi},n_a, n_b)(\theta_m)$ in the next step. For example, if it was a $f_a$ which emitted, the count $n_a$ will go to $n_a-1$. This means in particular that in the last sub-operation one has
\begin{align}
P(n_{\phi},n_a,n_b)(\theta_m) = P_{p_k}(\theta_m)
\,,
\end{align}
such that the last of the $2\times2$ sub-matrix is always of form
\begin{align}
U^{(m,m)}_h = \left( \begin{array}{cc}   0 & -1 \\ 1 & 0 \end{array} \right)
\end{align}
As a result, in the last sub-operation the amplitude of the $\ket{0}$ state of $\ket{h}_m$ is fully transferred to the $\ket{m}$ state. 

In the history register, this operation generates a superposition of states corresponding to all the possible emissions which could have happened. The amplitude for the emission to be associated with a particle $p_k$ is given by
\begin{align}
A_{p_k} = \sqrt{\frac{P_{p_k}}{\sum_{p_k} P_{p_k}}}
\,,
\end{align}
but this procedure includes the interference from all possible flavours each particle can have. 

Notice that at the end of the step, the  $U_-$ gates have been applied conditionally on all of the particles in $\ket{p}$, which is exactly the inverse of the first operation, where we counted the particles. As a result the three count registers will be back to the initial state $\ket{0}$ at the end of each step, ready to be used again. Furthermore, the emission register is reset back to $\ket{0}$ by the last controlled $ X$ operation. 

\subsection{Adjusting the particle state}
\label{sec:adjusting}
The final operation in each step adjusts the particle flavours according to the emissions that happened. For example, if a boson splits into a $f_a\bar{f}_a$ pair, we must remove a $\phi$ from $\ket{p}$ and add an $f_a$ and an $\bar{f}_a$. In general, if it is a fermion that emits we simply have to add a boson to the $\ket{p}$ register, while if it is a boson which emits we must add fermion-antifermion pair to $\ket{p}$ and remove the boson which emitted. The schematic of the circuit which performs this operation is given by
\[
\Qcircuit @C=1em @R=.5em @!R{
\lstick{\ket{p}_{M+1}} & {/} \qw & \sgate{U_p}{2} & \qw  \\
&	&    	 &\\
\lstick{\ket{p}_k} & {/} \qw & \gate{U_p} & \qw \\ 
&	&    	 &\\
\lstick{\ket{h}_m} & {/} \qw & \measure{\mbox{$k$}} \qwx[-2] & \qw  \\
}
\]

\vspace{-1mm}
\noindent As before, $M = m + n_I $. Notice that if we control on $\ket{h}$ being in the $\ket{k}$ state, we apply $U_p$ to the $k^\text{th}$ sub-register $\ket{p}_k$ and the $(M+1)^\text{th}$ sub-register $\ket{p}_{M+1}$.

Each sub-register in $\ket{p}$, made up of three qubits, corresponds to one particle state and can be in any of six possible states: $0$, $\phi$, $f_a$, $f_b$, $\bar{f}_a$ and $\bar{f}_b$. The sub-register $\ket{p}_{M+1}$ will encode the new particle which has just been emitted and it always starts out in the $0$ state, while the registers below encode the previous particle states. The operation labeled $U_p$, conditional on the $k^\text{th}$ state in $\ket{h}$, is a map from the $k^\text{th}$ and $(M+1)^\text{th}$ particle states before the emission, and the same particle states after the emission. Notice that this operation is controlled on the history states, which specify which particle emitted, though they do not hold the information of what kind of particle that was. That information is provided by the $k^\text{th}$ particle state. 
The $U_p$ gate is always the same and we want it to take
\begin{align}
\ket{f_i}\ket{0} & \rightarrow \ket{f_i}\ket{\phi} \nonumber \\
\ket{\bar f_i}\ket{0} & \rightarrow \ket{\bar f_i}\ket{\phi} \nonumber \\
\ket{\phi}\ket{0}& \rightarrow \sum_{i = a,b}\hat g_i \,  \left( \ket{f_i}\ket{\bar f_i} + \ \ket{\bar f_i}\ket{f_i} \right)
\,,
\end{align}
where
\begin{align}
\hat g_i \equiv \frac{g_i}{\sqrt{2(g_a^2+g_b^2)}}
\,.\end{align}
Here we used the vector representation of the particle states given in Eq.~\eqref{eq:states}. We can write this transformation as a single unitary operator as follows:
\begin{align}
\label{eq:Up}
U_p = & \sum_{i=a,b} \ket{f_i}\ket{\phi}\bra{f_i}\bra{0} + \sum_{i=1,b} \ket{\bar f_i}\ket{\phi}\bra{\bar f_i}\bra{0} \\
& + \sum_{i=a,b} \hat g_i \,  \left(\ket{f_i}\ket{\bar f_i} + \ket{\bar f_i}\ket{f_i} \right)\bra{\phi}\bra{0}\nonumber
\,.\end{align}
Since the particle states of different flavours are orthogonal to one another, this transformation is unitary.  

\subsection{Circuit Decomposition}
\label{sec:circuitdecomposition}

We now explain in some detail how to break down the operations in our general quantum circuit (including $\phi\rightarrow f\bar{f}$ and the running coupling) into standard quantum gates (single qubit gates and CNOT gates), so that we can run the circuit on a simulator and eventually on an actual testbed. While every effort was made to find an efficient breakdown of the circuit, we anticipate that a reduction in the number of standard quantum gates is still possible. The following discussion gives the number of gates required for each step $0 \leq m < N-1$ in the evolution. Table~\ref{tab:complexity} gives the number of gates needed after summing over all steps. 

\subsection{The first sub-operation, $U_\text{count}$}
We start with the counting operation. We store integers in the counting registers using the conventional bit representation, then the $U_+$ gate can be implemented as shown below.
\[
\Qcircuit @C=.6em @R=.6em @!R{ 
\lstick{w_{\ell -1}} & \qw &  \qw & \qw & \qw & \qw & \qw &  \qw & \ldots & & \targ & \ctrl{10} & \targ & \qw & \ldots & & \qw & \qw & \qw    \\
\lstick{w_{\ell -2}} &  \qw &  \qw & \qw & \qw & \qw & \qw &  \qw & \ldots & & \ctrl{-1} & \qw & \ctrl{-1}&  \qw & \ldots & & \qw & \qw  & \qw   \\
&  & & &  \ldots & & & & \ldots & & & & & &  \ldots  & &  &  & \\
\lstick{w_2} &  \qw & \qw & \qw & \qw & \targ & \ctrl{4} &  \qw & \ldots & & \qw & \qw & \qw &  \qw & \ldots  & & \targ & \qw  \\
\lstick{w_1} &  \qw & \qw & \targ & \ctrl{2} & \ctrl{-1} & \qw &  \qw & \ldots & & \qw & \qw & \qw &  \qw & \ldots  & & \ctrl{-1} & \targ & \qw  \\
\lstick{q_1} & \qw & \gate{X} & \ctrlo{-1} & \qw & \qw & \qw &  \qw & \ldots & & \qw & \qw & \qw &  \qw & \ldots  & & \qw &  \ctrlo{-1} & \qw   \\
\lstick{q_2} &\qw & \qw & \qw & \targ & \ctrlo{-2} & \qw & \qw & \ldots  & & \qw & \qw & \qw &  \qw & \ldots & & \ctrlo{-2} & \qw   & \qw  \\
\lstick{q_3} &  \qw & \qw & \qw & \qw & \qw & \targ & \qw & \dots &  & \qw & \qw & \qw &  \qw & \ldots   & & \qw & \qw & \qw    \\ 
&   &  & & \ldots & & & & \ldots  & & & & & & \ldots & & & & \\
 \lstick{q_{\ell -1}} &\qw &  \qw & \qw & \qw & \qw & \qw & \qw &  \ldots & & \ctrlo{-8} & \qw & \ctrlo{-8} &  \qw & \ldots   & & \qw & \qw   & \qw   \\
\lstick{q_\ell } & \qw \qw & \qw & \qw & \qw & \qw & \qw & \qw &  \ldots &  & \qw & \targ & \qw & \qw &  \ldots & & \qw & \qw & \qw   
}
\]

A general integer $a$ has the form $\ket{q_\ell ...q_3q_2q_1}$, where $\ell  = \lceil \log_2(a) \rceil$ is the number of bits necessary to store the integer (we round up to the nearest integer). Therefore, in our circuit the number of gates needed to implement a specific $U_+$ gate depends on the maximum integer we might have to store. In our circuit, the $U_+$ gate is controlled on the particle state $\ket{p}$ being a type $a$ fermion, a type $b$ fermion or a $\phi$ boson. From the definition of the particle state in Eq.~\eqref{eq:states} one can see that the first two cases require controlling on two qubits from $\ket{p}$, while the latter case requires controlling on all three the qubits from $\ket{p}$. These controls are applied to all of the operations of the $U_+$ decomposition above, yielding many instances of an $X$-gate being controlled on multiple qubits. It is a known result (see e.g. Ref.~~\cite{Nielsen:2011:QCQ:1972505}) how to decompose a $C^{(n)}(U)$ operation, requiring $n-1$ work qubits, $2 \times (n-1)$ Toffoli gates plus a $C(U)$ operation. A Toffoli gate requires 16 standard gates while a $C(U)$ operation where $U$ is real requires 5 standard gates in general (although if $U=X$ it is simply a CNOT gate). For $n>2$ and controlling on all qubits being in the $\ket{1}$ state, we then need 
\begin{align}
\left|C^{(n)}[X]\right| &= 32 n- 31
\nonumber\\
\left|C^{(n)}[U]\right| &= 32 n- 27
\end{align}
standard gates. To this count we add 2 $X$-gates for each time we control on a qubit being in the $\ket{0}$ state instead of the $\ket{1}$ state. Using these results, the total number of standard gates necessary for the counting operation when simulating the $m^\text{th}$ step is:
\begin{align}
909\lceil\log_2(m+n_I)\rceil- 1010
\,.
\label{eq:NCount}
\end{align}
The above number includes many pairs of adjacent $X$ gates (coming from controlling on a $\ket{0}$, rather than $\ket{1}$) that cancel. Ignoring all such $X$ gates gives 
\begin{align}
\label{eq:cCount}
c_{\rm count}(m, n_I) = 873\lceil\log_2(m+n_I)\rceil- 968
\,.
\end{align}
The true answer lies in between \eqref{eq:NCount} and \eqref{eq:cCount}; the effect is small and henceforth we ignore the difference arising from controlling on $\ket{0}$ versus $\ket{1}$. We therefore write the final answer as
\begin{align}
\label{eq:step1Count}
N_{\rm sub 1}(m, n_I) = c_{\rm count}(m, n_I)
\,.
\end{align}

\subsection{The second sub-operation, $U_e^{(m)}$}
Let's now look at the operation in which we determine whether or not we had an emission, whose circuit is given by
\[
\Qcircuit @C=1em @R=.5em @!R{
\lstick{\ket{e}} & {/} \qw &  \gate{U^{(m)}_e} & \qw   \\
\lstick{\ket{n_{\phi}}} & {/} \qw & \measure{\mbox{$n_\phi$}} \qwx[-1]   & \qw  \\
\lstick{\ket{n_a}} & {/} \qw  &   \measure{\mbox{$n_a$}} \qwx[-1]   & \qw     \\ 
\lstick{\ket{n_b}} & {/} \qw &   \measure{\mbox{$n_b$}} \qwx[-1]  & \qw  
}
\]
If we are at the $m^\text{th}$ step, the largest number of particles we can have is $m+n_I$, while the minimum is $n_I$. This means that we have to apply $U_e$ gates controlled on all the possible combinations of three integers, ranging from $0$ to $m+n_I$, whose sum is in the range $[n_I, m+n_I]$. There are 
\begin{align}
c(m,n_I)=\frac{m+1}{6}(m^2+3\,m \, n_I+5m+3n_I^2+9n_I+6)
\end{align}
such such combinations. For each of these we run a $ C^{(3\lceil\log_2(m+n_I)\rceil}(U_e) $ operation, where the $U_e$ gates are $R_Y(\theta)$ rotations. Using the results from above about $C^{(n)}(U)$ operations, the total number of standard gates necessary for the emission operation is 
\begin{align}
\label{eq:step2Count}
N_{\rm sub 2}(m, n_I) = c(m,n_I) \left(96\lceil\log_2(m+n_I)\rceil - 27\right)
\,.
\end{align}

\subsection{The third sub-operation, $U_h$}
The next operation we need to break down is the creation of the emission history. If we are in the $m^\text{th}$ step of the evolution, we can have up to $m+n_I$ particles in $\ket{p}$, so we must run $m+n_I$ of the sub-operations discussed before. We notice that the second part of the circuit for the sub-operation is the same as the counting operation, except we have $U_-$ gates instead of $U_+$ gates. The $U_-$ gate is implemented very similarly to the $U_+$ gate, the only difference being that we control on work qubits being in the $\ket{1}$ state instead of the $\ket{0}$ state. Therefore, the number of standard gates necessary to apply the controlled-$U_-$ operations when simulating the $m^\text{th}$ step is $c_{\rm count}(m, n_I)$ given in Eq.\eqref{eq:cCount}.

In the first part of the sub-operation circuit, the gate $U_h$ has the same controls on the count registers that we found in the emission operation, plus having controls on the particle state and on the emission qubit (which has to be in the $\ket{1}$ state for an emission to have happened). As we have seen the particle state can be in any of six states specified by three qubits; however, the emission probability is the same for particle and anti-particle, and $U_h$ is the identity if the particle is in state $\ket{0}$. Therefore the number of possible combinations of particle state and count states in sub-operation $j$ is $3 \, c(m, n_I)$, which is the number of times we must apply $U_h$ gates controlled on $4+3\lceil\log_2(m+n_I-j)\rceil$ qubits. The history register contains states labeled by integers from $0$ to $m+n_I$ and we use the standard bit representation to encode these integers in qubits. For the $j^{th}$ sub-operation the matrix $U_h$ is an $a \times a$ unitary matrix, where $a = 2^b$ with
\begin{align}
b=\lceil\log_2(j)\rceil
\,.
\end{align}
 $U_h$ only differs from the identity matrix in the row / column $1$ and $(j+1)$, which form the submatrix in Eq.~\eqref{eq:Uh}. Therefore, $U_h$ is a particular type of two-level unitary transformation acting on $b$ qubits, which we call $U[b]$. There is a standard procedure to break such matrices $U[b]$ down into standard qubit gates, and one can use this result to derive a break-down of a controlled $U[b]$ transformation. One finds
 \begin{align}
\left| C^{(n)}[U[b]]\right| &= 32(n-1) + \left| C^{(1)}[U[b]]\right| \nonumber\\
 &= 32(n-1) + 2(b-1) \left| C^{(1)}[X]\right| + \left| C^{(1)}[U]\right| \nonumber\\
 &= 64 b^2 - 94 b + 32 n+3
 \,.
 \end{align}

For our case we have $n = 4+3\lceil\log_2(m+n_I-j)\rceil)$, and combining these results the total number of standard gates necessary to implement the controlled-$U_h$ operations is
\begin{align}
\label{eq:step3Count}
&N_{\rm sub 3}(m, n_I) =  \sum_{j=1}^{m+n_I} \Bigg[ c_{\rm count}(m, n_I)\nonumber \\
&  \qquad +  3 c(m, n_I) \left| C^{(4+3\lceil\log_2(m+n_I-j)\rceil)}[U[b]]\right|\Bigg]
\,.
\end{align}

\subsection{The fourth sub-operation, $U_p^{(m)}$}
Given the transformation in Eq. \eqref{eq:Up}, we can implement $U_p$ efficiently as 
\[
\Qcircuit @C=1.5em @R=1.2em @!R{
\lstick{} & \targ  & \qw & \qw  & \qw   & \gate{U_r} & \qw & \ctrlo{2}  & \qw   \\
\lstick{} &  \qw    & \qw & \qw & \gate{H}   & \qw & \ctrlo{1} &  \qw & \qw   \\ 
\lstick{} & \qw  & \targ & \ctrl{3}  & \ctrl{-1}   & \ctrl{-2} & \ctrl{2} &   \ctrl{1} &  \qw  
 \inputgroupv{1}{3}{.3em}{3em}{\ket{p_{k}}}\\
\lstick{} &  \qw & \ctrl{-1} & \qw & \qw   & \qw  &  \qw & \targ &  \qw  \\
\lstick{} & \qw & \ctrlo{-1} & \qw & \qw & \qw & \targ &  \qw & \qw \\ 
\lstick{} &   \ctrl{-5}  & \ctrlo{-1}  & \targ & \qw  & \qw & \qw &  \qw & \qw 
 \inputgroupv{4}{6}{.3em}{3em}{\ket{p_j}}
}
\]

\vspace{-1mm}
\noindent where $k=m+n_I$.
In the circuit $H$ is the Hadamard gate and $U_r$ is given by
\begin{align}
\label{eq:Ur}
U_r = \frac{1}{\sqrt{g_a^2+g_b^2}}\left( \begin{array}{cc}   g_a & -g_b \\ g_b & g_a  \end{array} \right).
\end{align}
The operation $U_p$ is controlled on the possible states in $\ket{h}$. There are $m+n_I$ such states, each requiring $\lceil \log_2(m+n_I) \rceil$ controls. Thus, for each of the $m+n_I$ occurrences of $U_p$ one adds  $\lceil \log_2(m+n_I) \rceil$ controls to each operation above. This gives
\begin{align}
\label{eq:step4Count}
N_{\rm sub 4}(m, n_I) = (m+n_I)\left(224 \lceil \log_2(m+n_I) \rceil + 143\right)
\,.
\end{align}
standard gates.

\subsection{Summary}
\label{sec:decompositonSummary}
Adding all sub-operations together and summing over $0 < m < N-1$, one finds that the overall scaling of our circuit is $N^5 \ln N$. Fig.~2 shows the number of gates as a function of $N$ for $N < 50$. 

Note that one can obtain a much shallower circuit requiring less qubits if one takes into account that in the end states with different history registers do not interfere with one another. This implies that one can measure the history register after the third operation in each step, and reset it back to zero. This collapses the quantum state to one with a definite history. Having a state with a definite history gives definite knowledge about the number of bosons $n_\phi$, as well as the total number of particles $n_{\rm tot}$. This is because the history allows us to infer how many emissions have happened, which means that the state has a definite number of particles; since one also knows at any step at which an emission happened if the emitting particle was a fermion or a boson, one knows the total number of bosons. Thus, instead of counting and keeping track of the 3 values $n_\phi$, $n_a$ and $n_b$, it suffices to only keep track of $n_a$ and from that derive $n_b = n_{\rm tot} - n_\phi - n_a$. Following similar steps outlined in this section in this case, one can easily see that the scaling of the depth of the circuit is reduced significantly, with an overall scaling of $N n_f^2 \ln n_f$, instead of the $N^5 \ln N$. Here $N$ denotes the number of steps as before, while $n_f$ is the total number of fermions in the event, which can be significantly less than $N$, depending on the size of the couplings $g_{1,2}$ and how many fermions were in the initial state. While current quantum hardware does not allow for such repeated measurements, this might be possible in the future. 

%

\subsection{Circuit with no $\phi\rightarrow f\bar{f}$}
\label{sec:nosplitting}

Ignoring the $\phi \rightarrow f\bar{f}$ splittings, ignoring the running coupling, and starting with only one fermion (possibly in a superposition) as the initial state, allows us to drastically simplify our quantum circuit, since all one needs now is a single qubit which represents the fermion flavours, and a boson register, which keeps track of whether or not a boson was emitted at a given step. This boson register is the equivalent to the emission register plus the particle register in the general circuit. We no longer need a history register, since we know the fermion is the only particle which can emit, nor do we need the count registers since in this limit the probability of a boson being emitted only depends on the flavour of the fermion. The full evolution can be carried out with the much simpler circuit
\[
\Qcircuit @C=0.8em @R=.7em @!R{
\lstick{\ket{\phi_{N}}}	& \qw		&  \qw			&\qw& \qw 		&\qw &\qw & \gate{U^a_n} & \gate{U^b_n}	& \qw			& \meter	\\
\lstick{ \ldots}		&			&  				&& 			& 			& 	 		&  		&&		& \ldots	\\
\lstick{\ket{\phi_1}}	& \qw		&  \gate{U^a_1}	& \gate{U^b_1} &\qw &\dots & &\qw 		& \qw		& \qw 			& \meter 	\\
\lstick{\ket{f}}		& \gate{U}	&  \ctrlo{-1}		& \ctrl{-1} 	&\qw & \dots&	& \ctrlo{-3} 	& \ctrl{-3}		& \gate{U^\dagger} & \meter 	\\
}
\]

The $U$ and $U^\dagger$ gates are the same as in Eq.~\eqref{eq:UfDef}, while the $U_i^{a/b}$ gates are given by the matrices
\begin{align}
\label{eq:ULRiDef}
U^{a/b}_k = \left( \begin{array}{cc} \sqrt{\Delta_{a/b}(\theta_k)} & -\sqrt{1-\Delta_{a/b}(\theta_k)} \\ \sqrt{1-\Delta_{a/b}(\theta_k)} & \sqrt{\Delta_{a/b}(\theta_k)} \end{array} \right)
\,,
\end{align}
which encode the amplitude for the fermion to emit or not emit a boson at a given step. These gates are controlled on the fermion state since the gate parameters depend on the flavour of the fermion. The circuit construction demonstrates that the scaling for generating a single event is linear with the number of steps.


We now discuss how to implement this circuit on currently available hardware, breaking down the controlled operations into standard gates, namely single qubit gates and CNOT gates. To achieve this, one first uses the well known result
\[
\Qcircuit @C=.8em @R=.1em @!R {
& \ctrlo{2} & \qw &  & & \gate{X} & \ctrl{2} & \gate{X} & \qw \\
& & & \push{\rule{.3em}{0em}=\rule{.3em}{0em}} & &&  \\
& \gate{U} & \qw & & & \qw & \gate{U} & \qw & \qw 
}
\] 
In our case the gate $U$ consists of a $R_Y(\theta)$ rotation gate. Furthermore, we use the fact that for an arbitrary controlled-$U$ operation, one has
\[
\Qcircuit @C=.8em @R=.1em @!R {
& \ctrl{2} & \qw & & & \qw & \ctrl{2} & \qw & \ctrl{2} & \qw & \gate{P} & \qw \\
& & & \push{\rule{.3em}{0em}=\rule{.3em}{0em}} & & & & & & & \\
& \gate{U} & \qw & & & \gate{C} & \targ & \gate{B} & \targ & \gate{A} & \qw & \qw
}
\]
where
\begin{align}
P = \left( \begin{array}{cc} 1 & 0 \\ 0 & e^{i\psi} \end{array} \right) 
\,,
\end{align}
and the following conditions are satisfied
\begin{align}
U = \exp(i\psi)AXBXC\ ;\ ABC=I
\,.
\end{align}
To apply this to the controlled-$R_Y(\theta)$ gate one chooses
\begin{align}
A=R_Y(\alpha)\ \ \ B=R_Y(\beta)\ \ \ C=R_Y(\alpha)
\,,
\end{align}
where $\alpha$, $\beta$ and $\psi$ satisfy
\begin{align}
\alpha = \frac{\theta}{4}\, \qquad \beta =- \frac{\theta}{2}\,, \qquad  \psi=0
\,.
\end{align}
This gives gates $A$, $B$, $C$, $P$ (where $P$ is the trivial identity matrix) that satisfy all conditions. Using this information one finds that each step requires a total of 12 simple quantum gates (8 single qubit gates and four CNOT gates), and in addition two transformations are required at the beginning and end of the circuit which also consist of single qubit gates. Generating a single event therefore requires a total of
\begin{align}
n_{{\rm gates}} = 12 \, N + 2
\end{align}
single qubit and CNOT gates. 

\subsection{A classical algorithm and comparison to the quantum algorithm}

As already mentioned, a simulation of this algorithm on a classical computer scales exponentially with the number of time steps, rather than polynomially as in the case of a quantum computer. In this section we give some more details on a possible classical algorithm. There are several different ways in which a classical simulation can be done, and we choose here a method that is similar in spirit to the method used on the quantum computer, namely performing a change of basis in the fermion sector in which the evolution is diagonal, and rotating to the original basis after the full evolution has been performed. This can be implemented classically using the following steps
\begin{enumerate}
\item Perform a basis rotation of the initial state from the mass basis $f_{1/2}$ to the diagonal basis $f_{a/b}$
\item Randomly pick an initial state containing either one fermion $f_a$ or one $f_b$ based on the amplitudes of the initial state
\item Perform a normal shower evolution, which results in a particular history of emissions with a particular combination of fermion flavors
\item Find all possible fermion flavor combinations $f_{a / b}$ and compute their amplitudes given the chosen history of emissions
\item Performing the basis rotation, find the amplitudes for all possible fermion combinations $f_{1 / 2}$
\item Choose one of the fermion combinations based on their probabilities (squared amplitudes) and reweight the generated event to this probability
\end{enumerate}

The complexity of this algorithms can be estimated as follows: To compute a given amplitude requires a fixed number of computation for each step, and the number of amplitudes that need to be computed scales as $2^{n_f / 2}$ with the number of fermions in the event. Thus the classical scaling is 
\begin{align}
{\rm [classical]} \sim N \times 2^{n_f / 2}
\end{align}
As discussed previously, the scaling of the quantum algorithm with and without the repeated measurements is
\begin{align}
{\rm [quantum \, repeat]} &\sim N \times n_f^2 \ln n_f\\
{\rm [quantum \, no \, repeat]} &\sim N \times N^5 \ln N
\end{align}
Since the relative prefactor depends on how precisely one counts the number of classical computations, we omit it in a comparison between the classical and quantum algorithm. The quantum algorithm with repeated measurements of the history register beats the classical algorithm once the number of emitted fermions exceeds roughly 10. To compare the quantum algorithm without the repeated measurement to the quantum algorithm depends on the average number of fermions emitted, which depends on the coupling $g_{L/R}$.

\begin{figure}
\centering
\includegraphics[width=0.45\textwidth]{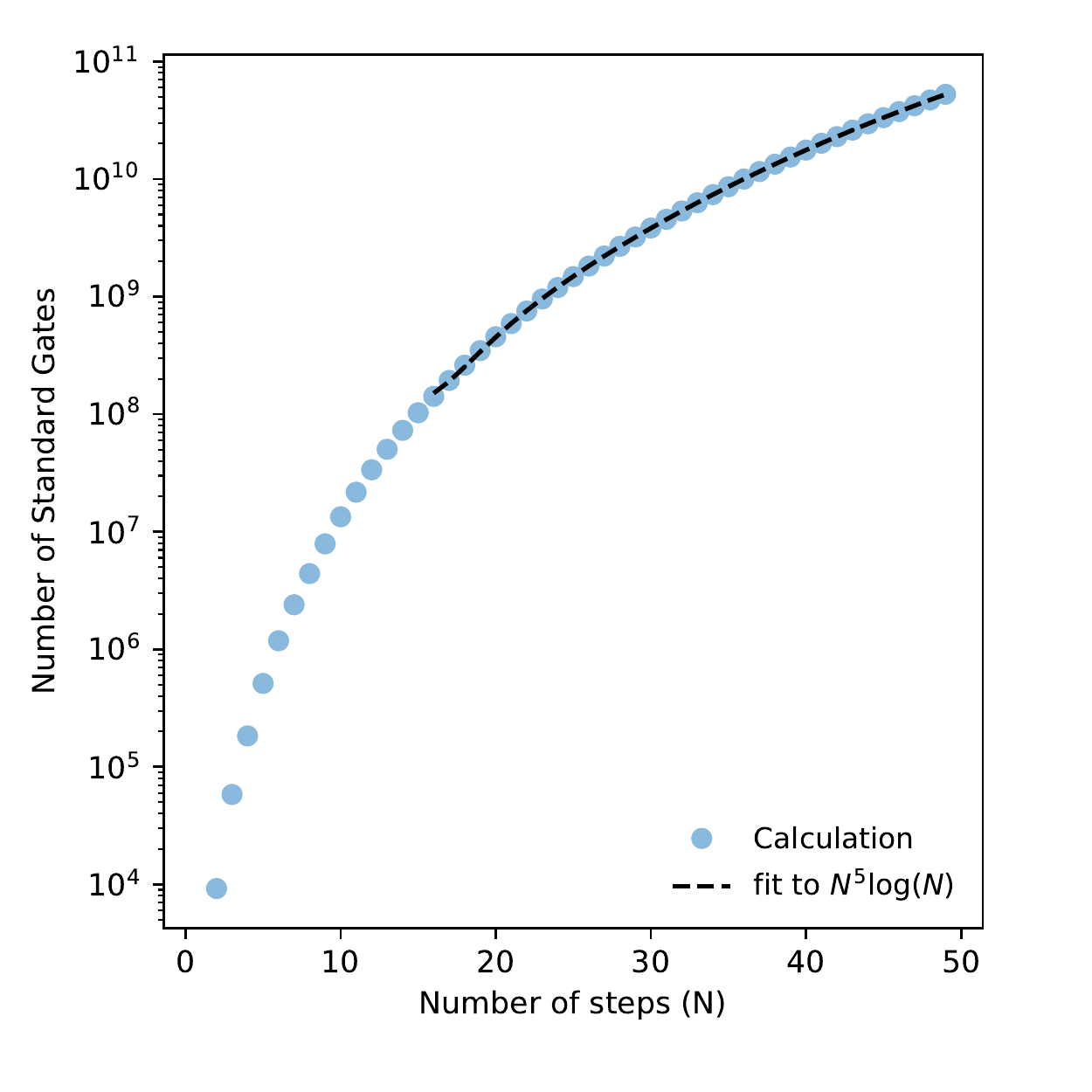}
\caption{The number of standard qubit gates as a function of the number of states, using the formulae given in Eqs.~\eqref{eq:step1Count},~\eqref{eq:step2Count},~\eqref{eq:step3Count} and~\eqref{eq:step4Count}
.  The asymptotic behavior is illustrated with a fit to $N^5 \ln N$.}
\label{fig:scalingplot}
\end{figure}

\subsection{Simulation experiments on IBM Q Johannesburg}
\label{sec:experiment}

\begin{figure}[h!]
\centering
\includegraphics[width=0.45\textwidth]{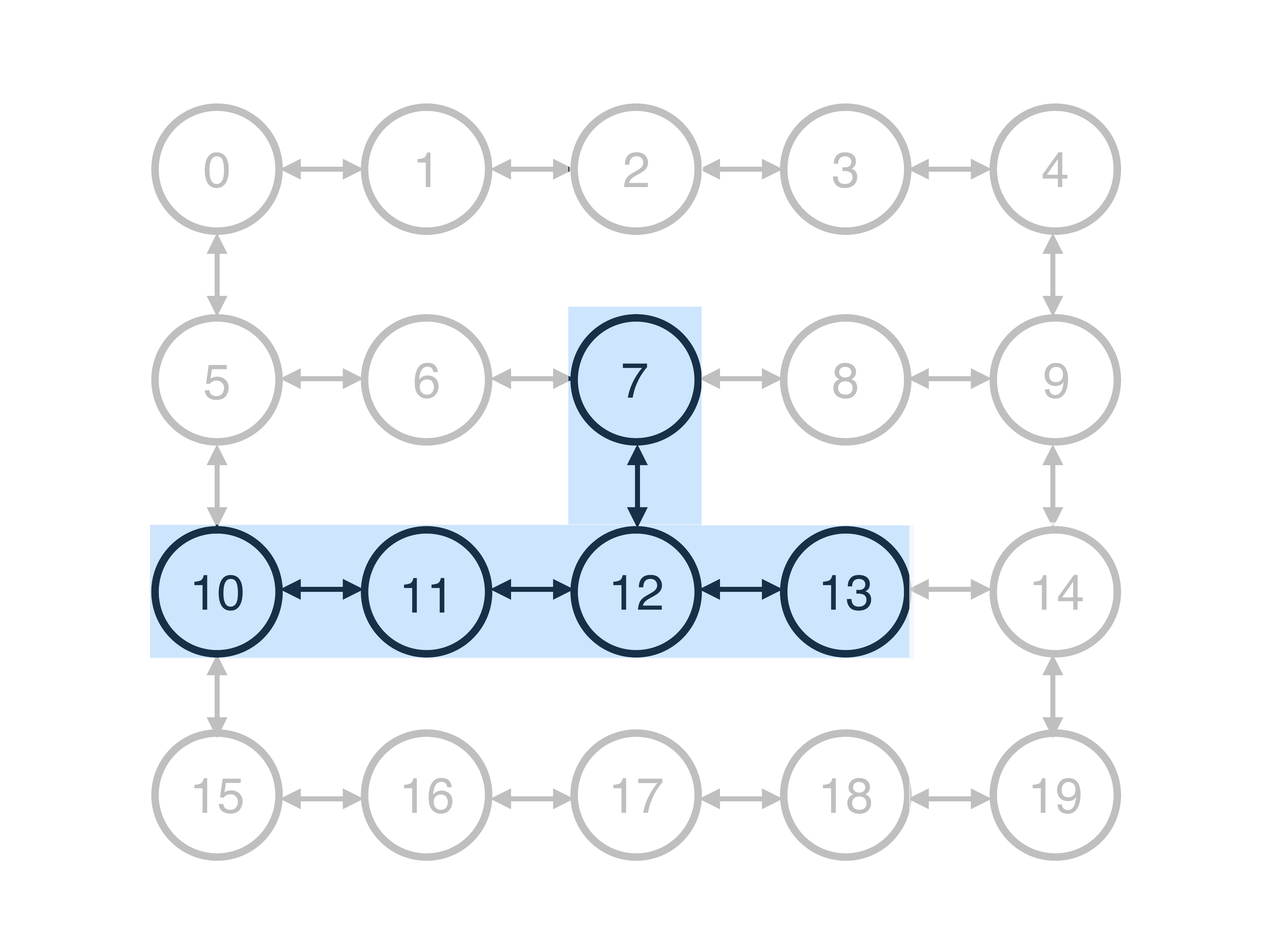}
\caption{Layout of IBM Q Hub Johannesburg chip. The highlighted qubits were used in generating the experimental data in Figure~\ref{fig:quantum} and were chosen to minimize the number of required SWAPs and operation errors.}
\label{fig:chip}
\end{figure}

In this section we will discuss the experimental quantum computer setup and data postprocessing for the 4 step simulation shown in Figure~\ref{fig:quantum} for the circuit without $\phi$ splitting. In our experiments we utilized the 20 qubit Johannesburg chip, available to members of the IBM Q Network. We choose qubit 12 as the \ket{f} qubit, which is connected to qubits 7, 13 and 11 (Figure~\ref{fig:chip}). In addition, we used qubit 10 as the fifth qubit. The choice of qubit was based on the best one- and two-qubit calibration data provided by IBM on the day experiments were run on the hardware.

The 4 step circuit requires 5 qubits with the \ket{f} qubit needing to perform two-qubit CNOT operations with the other 4 qubits in sequence. Our circuit consists of 53 gates, of which 34 are one-qubit rotations and 19 are two-qubit operations (of which 3 were needed to swap qubits 10 and 11). The maximum connectivity on the Johannesburg chip, and in fact all other available chips, on the IBM Q platform is 3. To connect the final qubit to the \ket{f} qubit, a minimum of one SWAP operation is required. Within the bit string the logical ordering of the qubits [0--4] is [12, 7, 13, 11, 10]. At the point in the circuit where qubits 12 and 10 need to interact, we simply apply a SWAP operation between qubits 10 and 11 and continue with the circuit operating between qubit 12 and 11. Instead of swapping qubits 10 and 11 back at the end of the operations, we simply read out the qubits in the expected order by assigning them to the correct classical registers containing the bit strings. The circuit was run 100 times with 8192 shots per run, resulting in a total of 819,200 samples.

\begin{figure}
\centering
\includegraphics[width=0.45\textwidth]{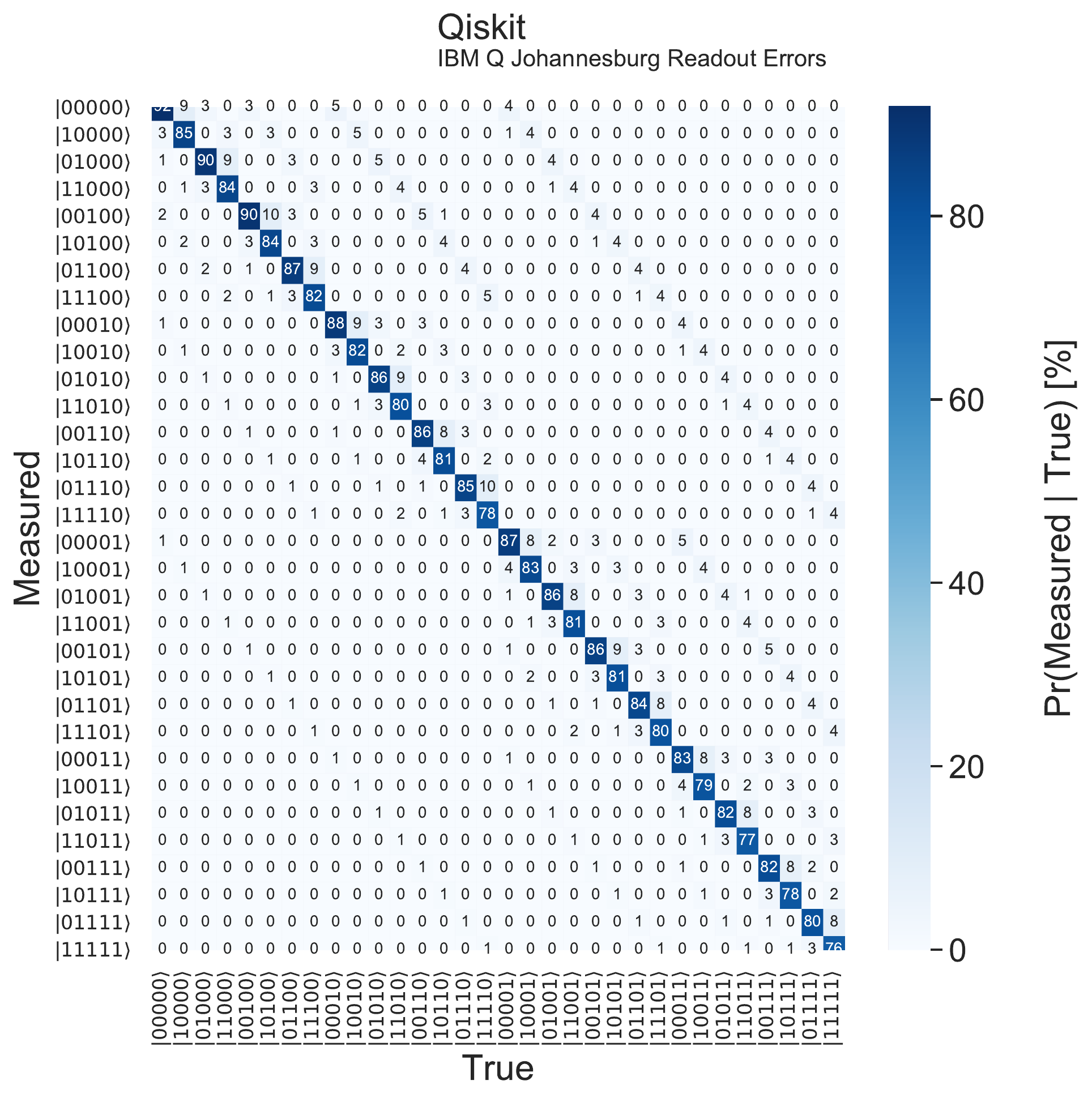}
\caption{The Johannesburg IBM Q response matrix used for readout error corrections with the mapping between qubits [0--4] and logical qubits described in the text.}
\label{fig:responsematrix}
\end{figure}

Readout noise is one of the largest sources of errors on a quantum computer. To correct for these errors, we used both IBM's constrained matrix inversion approach implemented in \texttt{qiskit-ignis}~~\cite{ignis2019} and the iterative Bayesian method described in Reference~~\cite{Nachman2019} with 100 iterations.  Both results are consistent with each other and the answer is insensitive to the number of iterations.  Applying either method requires a \textit{response matrix} encoding the migrations between qubit states before and after a measurement is performed.  The response matrix was generated immediately prior to running the circuits, within the same job.  For each of the $2^5$ possible states, a circuit is constructed to build that particular state by applying $X$ rotations.  This is repeated many times for each state and the outcomes populate the response matrix.  It should be noted that the response matrix will also accumulate errors from the 1-qubit rotation, though these are likely to be small.   The response matrix is presented in Fig.~\ref{fig:responsematrix}.  As desired, the most probable measured state is the original state for all cases.  Off diagonal terms account for 10-20\% of the probability mass function for each state.

In addition to the readout corrections, we made an attempt to mitigate the errors generated by the two-qubit CNOT operations. Here we followed the zero-noise extrapolation technique by Dumitrescu et al.~~\cite{Dumitrescu2018}, where each CNOT operation (with exception of the SWAP operation) was expanded respectively by 2 and 4 CNOT operations (adding effectively 1 or 2 identity operations). As our circuit has 16 CNOT operations, the extrapolation scheme required circuits with 48 and 80 CNOTs, respectively. To ensure that these extra gates were not removed, the circuits were executed setting the transpiler optimization level to 0.

\begin{figure*}
\centering
\includegraphics[width=0.5\textwidth]{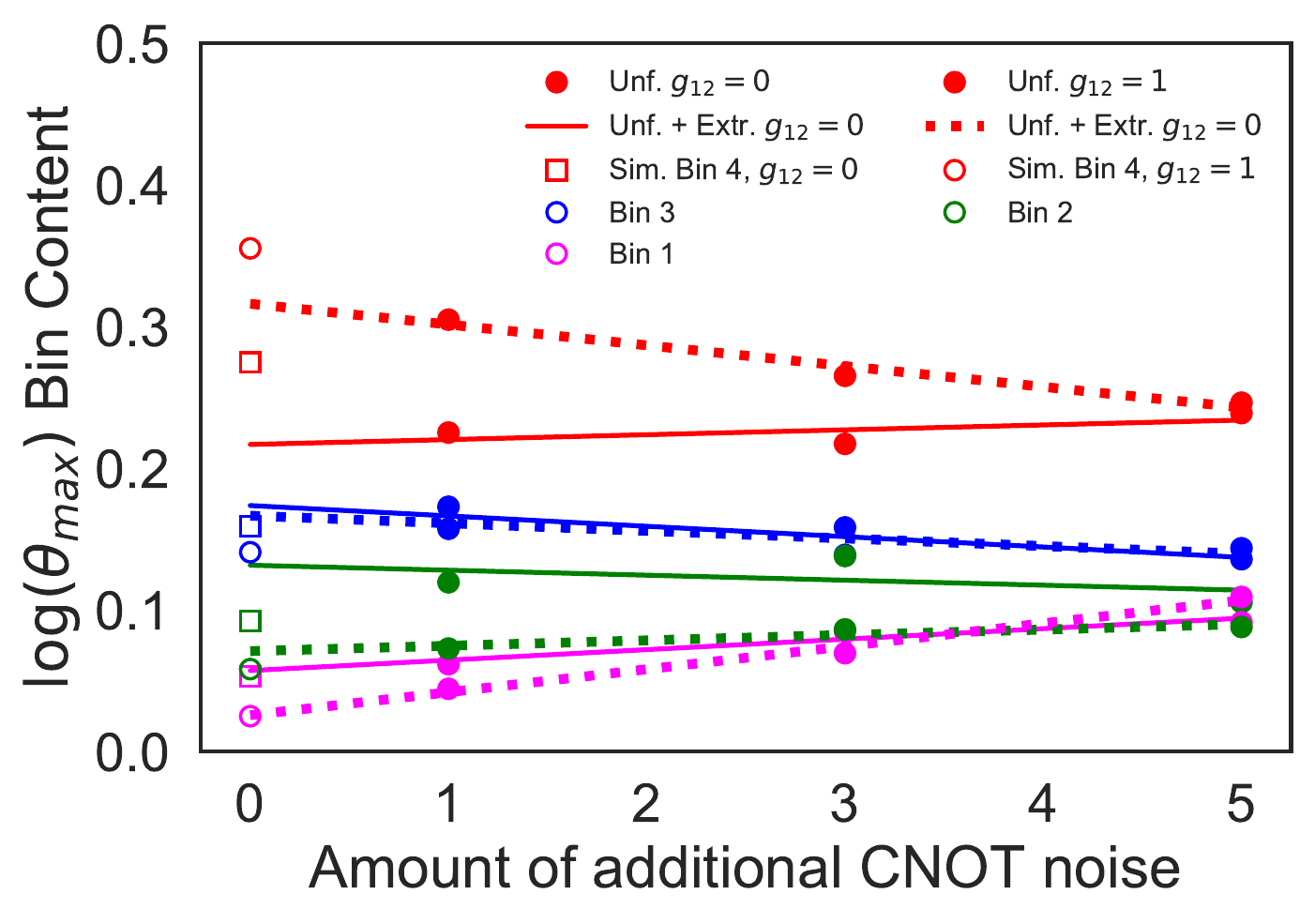}\includegraphics[width=0.5\textwidth]{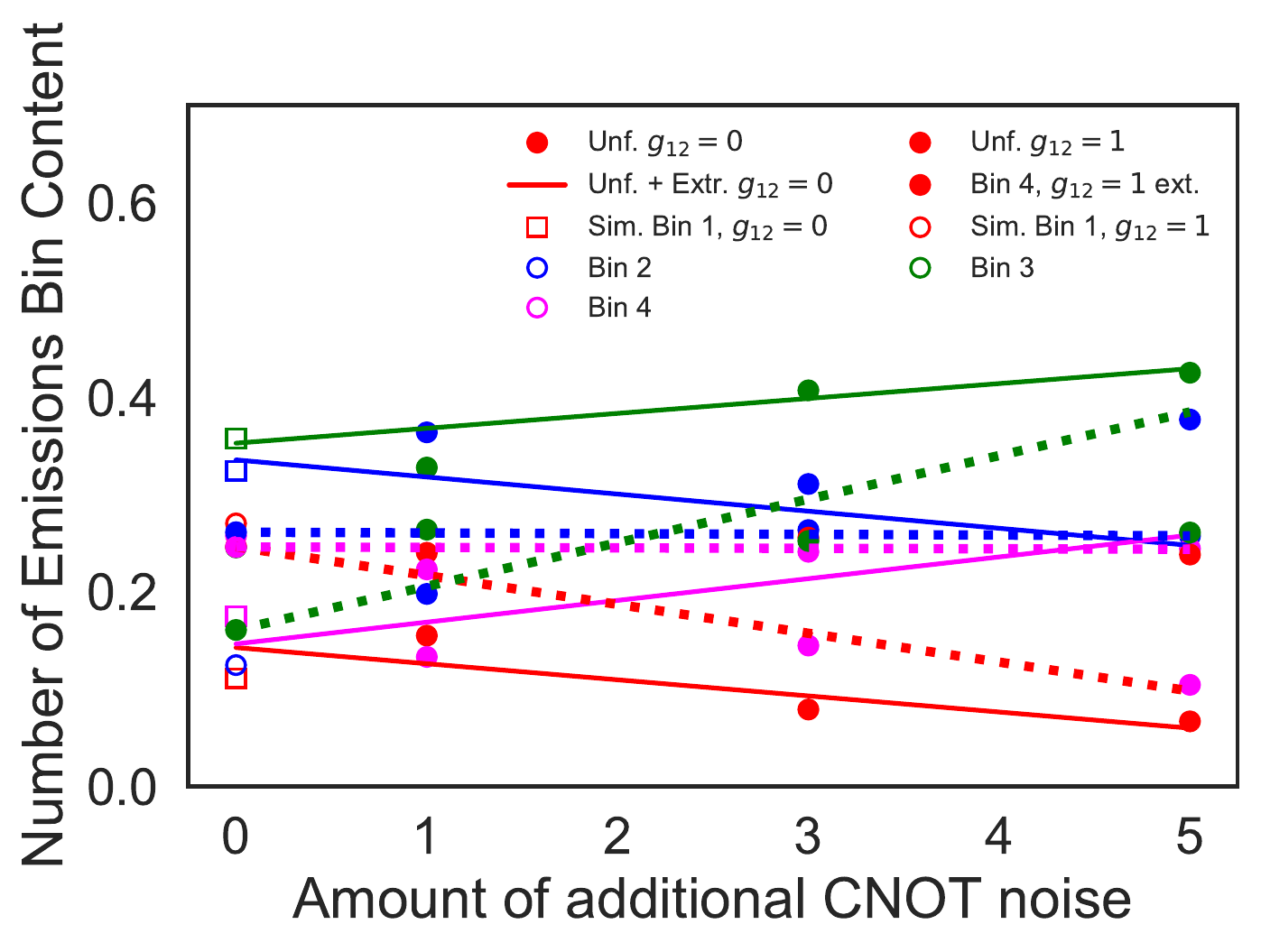}
\caption{An illustration of the CNOT extrapolation for $\log(\theta_\text{max})$ (left) and the number of emissions (right).  The lines show the result of a linear fit, the filled markers are the quantum measurements, and the open markers are the classical simulations of the quantum circuit.  Each bin is represented by a different color.}
\label{fig:extrapolation}
\end{figure*}

Modeling CNOT noise as simply additive, one can fit the value of any observable as a function of the number of CNOTs to a straight line and extrapolate to zero.  In our case, for each bin of the histograms in Figure~\ref{fig:quantum}, we have the bin value for $1$, $3$, and $5$ CNOTs.  A linear function is fit to these data and then evaluated at zero in order to extrapolate the CNOT errors to zero.  Figure~\ref{fig:extrapolation} presents the extrapolation for $\log(\theta_\text{max})$ and the number of emissions for both values of $g_{12}$.  The data seem to be approximately linear, but it is difficult to establish linearity with only three points.  As a result, the plots presented in Figure~\ref{fig:quantum} only have unfolding corrections.  Developing qubit and gate efficient methods for mitigating CNOT noise is an active area of research that may result in effective methods for further reducing the noise of this measurement in the future.

\begin{figure*}
\centering
\includegraphics[width=0.5\textwidth]{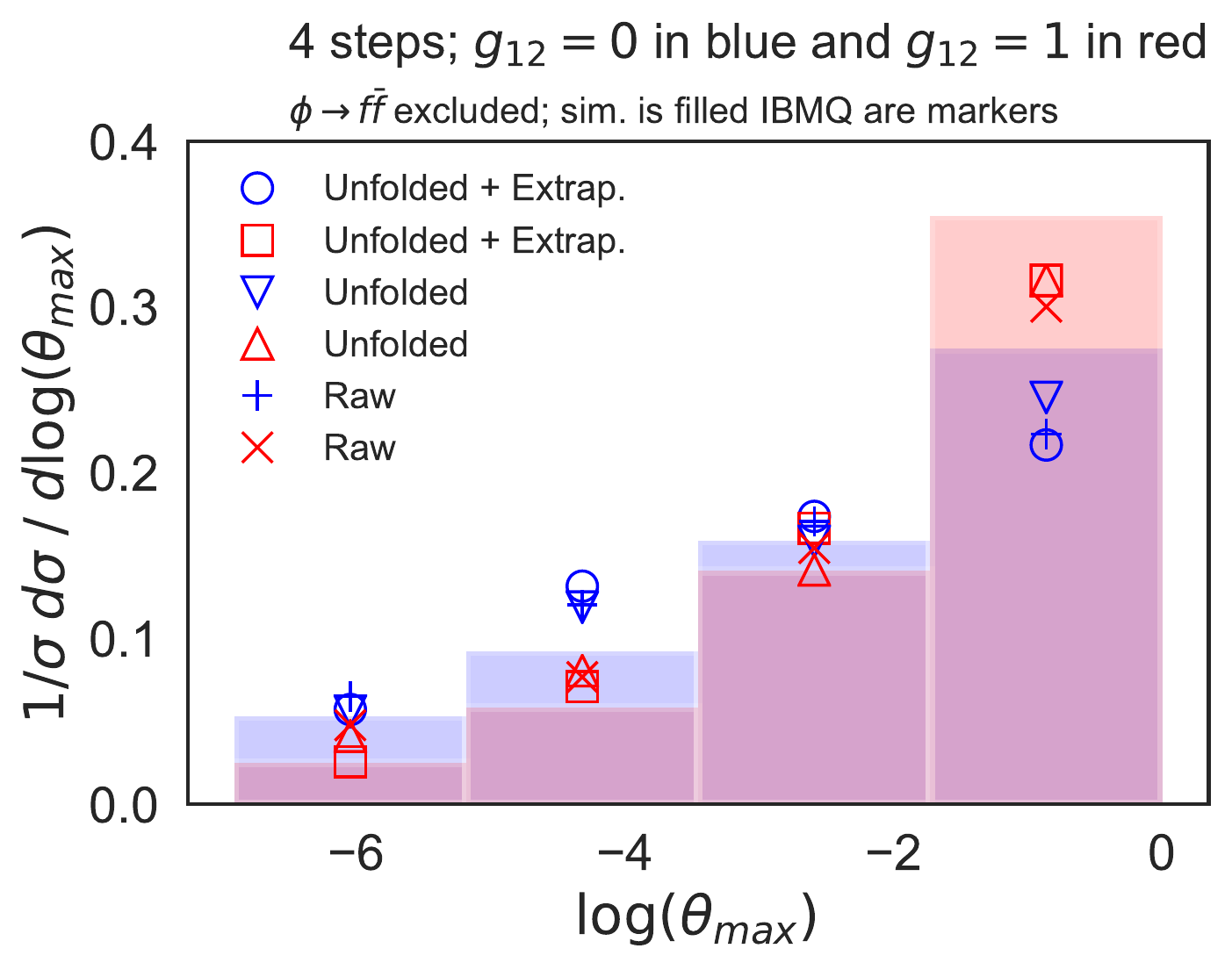}\includegraphics[width=0.5\textwidth]{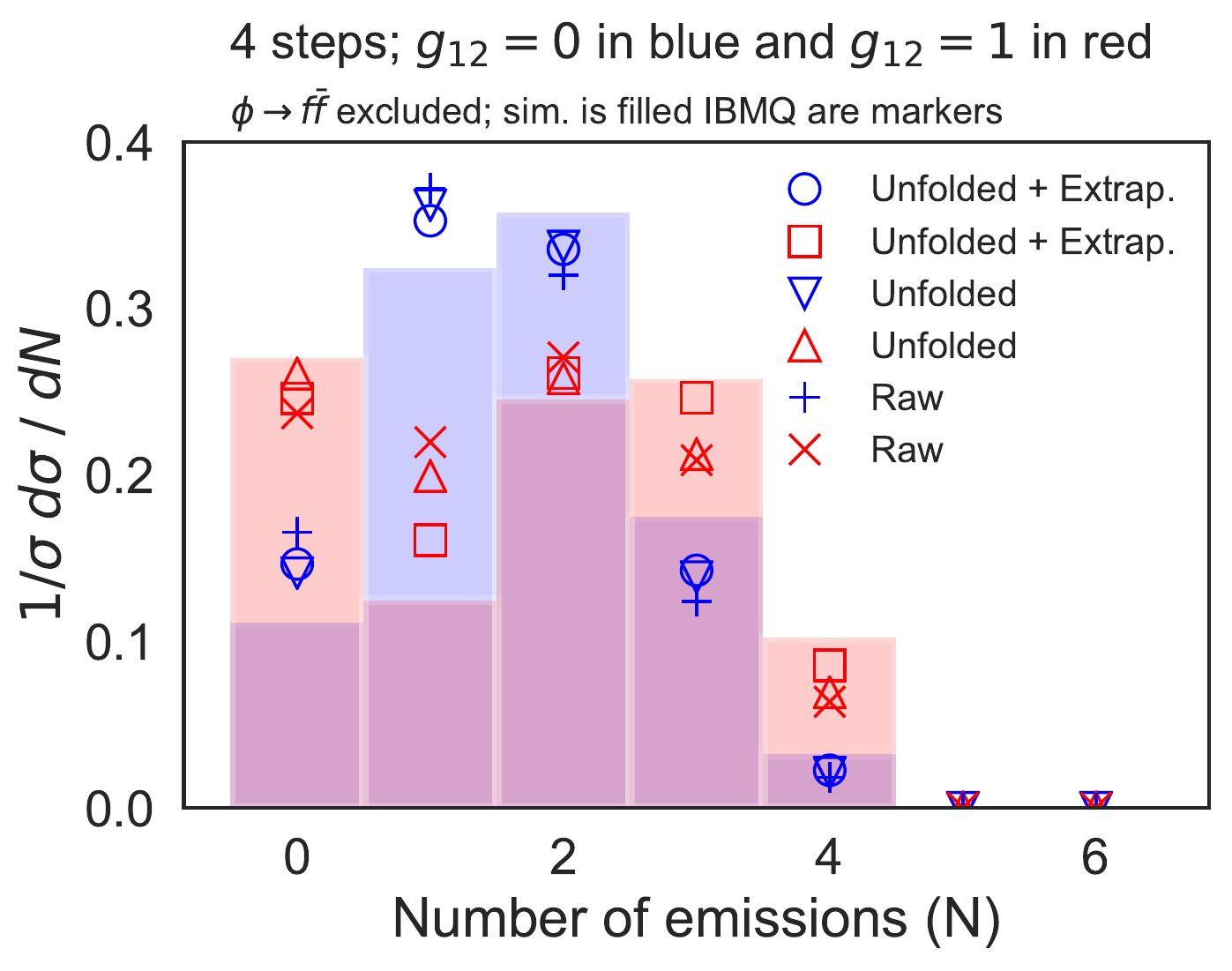}
\caption{The histograms of $\log(\theta_\text{max})$ (left) and the number of emissions (right) with the  measurements after various levels of corrections.}
\label{fig:correctionsummary}
\end{figure*}

A summary of the measurements and the various corrections are presented in Figure~\ref{fig:correctionsummary}.  The corrections are typically smaller than the difference between turning on and off interference effects and bring the measurements closer to the predictions.

\end{document}